\documentclass{aa}

\usepackage[varg]{txfonts}
\usepackage{amssymb}
\usepackage{epsfig}
\usepackage{graphics}
\usepackage{amsmath}
\usepackage{color}
\usepackage{natbib}
\usepackage{twoopt}
\usepackage{xtab}

\usepackage[switch, modulo]{lineno}

\newcommand{\be}{\begin{equation}}
\newcommand{\ee}{\end{equation}}

\newcommand{\fc}{\ensuremath{f_{\mathrm{c}}}}
\newcommand{\teff}{\ensuremath{T_{\mathrm{eff}}}}
\newcommand{\ledd}{\ensuremath{L_{\mathrm{Edd}}}}
\newcommand{\ke}{\ensuremath{\kappa_{\mathrm{e}}}}
\newcommand{\fedd}{\ensuremath{F_{\mathrm{Edd}}}}
\newcommand{\tedd}{\ensuremath{T_{\mathrm{Edd}}}}

\newcommand{\Zc}{\ensuremath{\mathbb{Z}}} 
\newcommand{\Ac}{\ensuremath{\mathbb{A}}} 

\begin{document}

\title{Models of neutron star atmospheres enriched with nuclear burning ashes}

\author{J.\,N\"attil\"a\inst{1}\thanks{joonas.a.nattila@utu.fi} 
\and V.\,F.\,Suleimanov\inst{2,3}
\and J.\,J.\,E. Kajava\inst{4}
\and J.\,Poutanen\inst{1}
}

\institute{Tuorla Observatory, Department of Physics and Astronomy, University of Turku, V\"ais\"al\"antie 20, FI-21500, Piikki\"o, Finland 
\and Institute for Astronomy and Astrophysics, Kepler Center for Astro and Particle Physics, Eberhard Karls University, Sand 1, 72076 T\"ubingen, Germany 
\and Kazan (Volga region) Federal University, Kremlevskaya 18, 420008 Kazan, Russia 
\and European Space Astronomy Centre (ESA/ESAC), Science Operations Department, 28691 Villanueva de la Ca\~nada, Madrid, Spain}

\date{Received 11/05/2015 / Accepted 25/06/2015}

\abstract{Low-mass X-ray binaries hosting neutron stars (NS) exhibit thermonuclear (type-I) X-ray bursts, which are powered by unstable nuclear burning of helium and/or hydrogen into heavier elements deep in the NS ``ocean''.
In some cases the burning ashes may rise from the burning depths up to the NS photosphere by convection, leading to the appearance of the metal absorption edges in the spectra, which then force the emergent X-ray burst spectra to shift toward lower energies.
}
{These effects may have a substantial impact on the color correction factor $f_{\mathrm{c}}$ and the dilution factor $w$, the parameters of the diluted blackbody model $F_{{E}} \approx w B_{{E}}(f_{\mathrm{c}} T_{\mathrm{eff}})$ that is commonly used to describe the emergent spectra from NSs.
The aim of this paper is to quantify how much the metal enrichment can change these factors.
}
{We have developed a new NS atmosphere modeling code, which has a few important improvements compared to our previous code required by inclusion of the metals.
The opacities and the internal partition functions (used in the ionization fraction calculations) are now taken into account for all atomic species.
In addition, the code is now parallelized to counter the increased computational load.
}
{We compute a detailed grid of atmosphere models with different exotic chemical compositions that mimic the presence of the burning ashes.
From the emerging model spectra we compute the color correction factors \fc\ and the dilution factors $w$ that can then be compared to the observations.
We find that the metals may change \fc\ by up to about $40\%$, which is enough to explain the scatter seen in the blackbody radius measurements.
}
{The presented models open up the possibility for determining NS mass and radii more accurately, and may also act as a tool to probe the nuclear burning mechanisms of X-ray bursts.
}


\maketitle

\section{Introduction}\label{sec:intro}

Low-mass X-ray binaries (LMXB) that consist of a neutron star (NS) primary, and a secondary star less massive than the Sun, may exhibit thermonuclear (type-I) X-ray bursts \citep[see e.g.][for review]{LvPT93, SB06}.
X-ray bursts are produced when the mass accretion rate onto the NS is in a regime where unstable nuclear fusion burns the accumulated hydrogen and/or helium into heavier elements \citep{WT76,FHM81}.
The heat generated in the burning is released as X-ray radiation from the NS photosphere \citep{Joss78}, and some X-ray bursts are so energetic that the luminosity exceeds the local Eddington limit.
In these cases the radiation pressure lifts the photosphere from the NS surface, and we can observe so-called photospheric radius expansion (PRE) bursts \citep{HCL80}.
These PRE bursts are particularly interesting as they can be used to constrain the NS mass and radius by comparing the observed cooling tracks to the predictions from theoretical NS atmosphere models \citep{SPW11, SPW12, PNK14}.

The nuclear burning region is a thin layer located deep in the neutron star ``ocean'' \citep{FHM81,FL87}, at column density on the order of $10^{8}\,{\rm g \, cm}^{-2}$ \citep{CB00}.
Detailed one-zone nucleosynthetic studies and hydrodynamical models coupled with vast reaction networks show that large quantities of heavier elements are produced during the nuclear burning process \citep[see e.g.][for a review]{PJSI13}.
In PRE bursts the conditions are favorable for large scale convection and the burning ashes may rise up to the photosphere \citep{WBS06}.
The presence of these heavier elements in the H and/or He plasma can then significantly alter the properties of the atmosphere.

Theoretical calculations show that the emergent energy spectrum emanating from the hot X-ray bursting NS photosphere is closely approximated by a diluted blackbody $F_{{E}} \approx w B_{{E}}(f_{\mathrm{c}} T_{\mathrm{eff}})$, where $w$ is the dilution factor, $B_{{E}}$ is the Planck function, $f_{\mathrm{c}} \equiv T_{\mathrm{c}}/T_{\mathrm{eff}}$ is the color correction factor, $T_{\mathrm{eff}}$ is the effective temperature, and $T_{\mathrm{c}}$ is the observed color temperature \citep{London84, SPW11, SPW12}.
If the burning ashes reach the NS photosphere, strong absorption edges may appear on top of the diluted blackbody spectrum.
The strength of the edges depends on the composition of the ashes, but also on the NS photospheric temperature \citep{WBS06}. 
Encouragingly, \cite{iZW10} have found evidence that several spectra of PRE bursts with very strong superexpansion show strong residuals that can be modeled as absorption edges of heavy elements like $^{56}$Ni.
However, this evidence is not conclusive with the current instruments, which have poor energy resolution.
The origin of the absorption edges are hard to verify, for example, the edges (and the emission lines) may also be produced by reflection from the surrounding accretion disk \citep[e.g.][]{SB02}.

Fortunately, the burning ashes have another effect on the X-ray burst spectra.
If the metals reach the photosphere, they will increase the bound-free and the free-free opacity, and also the photospheric electron density, causing the emergent spectra to shift toward lower energies.
In this case, the color correction factor $f_{\mathrm{c}}$ may decrease substantially.
Given that each X-ray burst is likely to ignite in slightly different initial conditions, the amount and the composition of the burning ash that reaches the photosphere is expected to be variable between different bursts.
Therefore, the burning ashes may also manifest themselves as small burst-to-burst variations in the cooling tracks of individual bursts because the observed blackbody radii have a strong dependency on the color correction factor ($R_{\rm bb} \propto \fc^{-2}$).
In LMXBs where the \textit{RXTE}/PCA have detected several PRE bursts from one source, one indeed sees such $R_{\rm bb}$-scatter \citep[e.g.][]{BMG10, ZMA11, GPO12, PNK14}. 
Even if there are other possible mechanisms that may be responsible for it (such as anisotropic burst emission), 
the fluctuations of \fc\  by the burning ashes is one of the strongest candidates.
Motivated by this possibility, we have undertaken a study to quantify the effects of these nuclear burning ashes on NS atmospheres and at the same time to improve our estimates of the NS masses and radii.


The paper is structured as follows.
In Sect. \ref{sec:method}, we describe the methods we use to model the atmospheres and the emergent spectra.
Here we present our new updated atmosphere code, originally based on the work by \citet{SPW11,SPW12}.
In order to validate our new methods, we compare our results to earlier work done by \citet{SPW12} and discuss the accuracy of the calculations.
Then, in Sect. \ref{sec:newgrid}, a new grid of models is presented for various chemical compositions consisting of a solar mixture of H and He with highly enhanced metallicities.
We also determine the color correction factors for the metal-enriched atmospheres and show how they can be applied to observed X-ray burst data. 
At the end of this section we discuss the impact these findings may have on the measurements of NS masses and radii and how they can act as a tool to probe the nuclear burning mechanisms.
Finally, we summarize the main findings in Sect. \ref{sec:summary}.

\section{Method of atmosphere modeling}\label{sec:method}

The methods we use to compute the NS atmosphere models share many important features with earlier works by \citet{SPW11} and by \citet{SPW12}.
For completeness, we present the main equations and basic assumptions in Sect. \ref{sect:theory}, giving emphasis to the improvements done for the level population and opacity calculations in Sect. \ref{sect:improvements}.
To solve these equations efficiently we have developed a new atmosphere modeling code \texttt{matmos}.
It has its origin in the old stellar atmosphere modeling program \texttt{ATLAS} \citep{K70, K93} which has been more recently modified to deal with high temperatures \citep{I03,SulP06, SW07} and to take into account Compton scattering \citep{SPW12} using exact relativistic redistribution functions \citep[see e.g.][]{PS96}.
From there on, the code was redesigned and rewritten using the modern, high-level computer programming language \texttt{julia} \citep{julia2012} to allow efficient and easy scaling from desktop computers to supercomputing clusters.
Parallelization is now done natively for high demand and heavy load routines like opacity and redistribution function calculations.

\subsection{Main equations}\label{sect:theory}

In the process of modeling the atmospheres, we have to set a few free parameters.
The first explicit input parameter in our model is the surface gravity
\be
\label{eq:gravity}
g\!=\!\frac{G M}{R^2}(1+z),
\ee
which depends on the mass $M$ and the radius $R$. 
Here $G$ is the gravitational constant and the term $1+z\!=\!(1-R_{\mathrm{S}}/R)^{-1/2}$ (where $R_{\mathrm{S}}\!=\!2 G M/c^2$ is the NS Schwarzschild radius 
and $c$ is the speed of light) is the general relativistic correction. 
In the computations one must also set the stellar effective temperature \teff\ of the NS.
However, a more convenient variable is the relative luminosity $\ell \equiv L/\ledd$, where \ledd\ is the Eddington luminosity measured at the surface.
It is common to define \ledd\ as
\be\label{eq:ledd}
\ledd = \frac{4\pi G M c}{\ke}(1+z),
\ee
where the Thomson scattering opacity $\ke$ is given through the Thomson cross-section $\sigma_{\mathrm{T}}$, the gas density $\rho$ and the electron number density $N_{\mathrm{e}}$ as
\be \label{eq:kappae}
\ke = \sigma_{\mathrm{T}} \frac{N_{\mathrm{e}}}{\rho} \approx 0.2(1+X)~\mathrm{cm}^2~\mathrm{g}^{-1} ,
\ee
where $X$ is the hydrogen mass fraction.
We note, however, that this is just a definition that we adopt to be consistent with previous work.
In reality, the electron cross-section is affected by the Klein-Nishina effect and the final form used for the Thomson scattering opacity $\ke$ is just an approximation where full ionization is presumed and the ratio between the number of electrons to protons and neutrons in the metals is assumed to be half.
Defining the Eddington flux and the Eddington temperature as 
\be \label{eq:fedd}
\fedd = \frac{\ledd}{4\pi R^2} = \sigma_{\mathrm{SB}} \tedd^4 = \frac{g c}{\ke},
\ee
where $\sigma_{\mathrm{SB}}$ is the Stefan-Boltzmann constant, and noting that $\ell = F/\fedd$,
we find a simple relation 
\be
\teff = \ell^{1/4} \tedd. 
\ee
It can be used to set the effective temperature of the atmosphere easily via the relative luminosity.
Lastly, we can freely set the chemical composition of our atmosphere.

The structure of the atmosphere is described by a set of differential equations that originate from our underlying assumptions and simplifications.
The first assumption is that the atmosphere is in the hydrostatic equilibrium
\be
\frac{\mathrm{d}P_\mathrm{g}}{\mathrm{d}m}\!=\!g-g_{\mathrm{rad}},
\label{eq:hydroeq}
\ee
where $P_{\mathrm{g}}$ is the gas pressure, $g_{\mathrm{rad}}$ is the radiative acceleration and the column density $m$ is found from
\be
\mathrm{d}m\!=\!-\rho \mathrm{d}z,
\ee
where $z$ the vertical distance.
This assumption is not valid in the radius expansion stage of PRE bursts and, therefore, our models are limited to their cooling phases (and to the non-PRE bursts).

The second important element in the theory is the radiative transfer equation (RTE).
In our calculations we assume radiative equilibrium meaning that radiation field is the only means of energy transport; i.e. both convection and thermal conduction are neglected.
We can simplify the RTE by assuming a plane-parallel atmosphere because the scale height of the atmosphere is smaller (on the order of $\sim \! 10$ to $\sim \! 10^3$ cm) than the NS radius ($\sim \! 10^6$ cm).
In the plane-parallel approximation the RTE can be formulated with the specific intensity $I(x,\mu)$ and the source function $S(x,\mu)$ as
\be\label{eq:RTE}
\mu \frac{\mathrm{d}I(x,\mu)}{\mathrm{d}\tau(x,\mu)} = 
I(x,\mu)-S(x,\mu),
\ee
where $\mu\!\equiv\! \cos \theta$ is the cosine of the angle between the surface normal and the direction of photon propagation. 
The dimensionless photon energy $x\!=\!h \nu / m_{\mathrm{e}} c^2$ is given in the units of electron rest mass energy $m_{\mathrm{e}}c^2$, where $h$ is the Planck constant, $\nu$ is the photon frequency and $m_{\mathrm{e}}$ is the electron rest mass.
The optical depth can be related to the opacity and the column density $m$ by
\be
\mathrm{d}\tau(x,\mu)\!=\![\sigma(x,\mu)+k(x,\mu)]\mathrm{d}m,
\ee
where $k(x,\mu)$ is the true opacity due to free-free and bound-free transitions and $\sigma(x,\mu)$ is the electron scattering opacity.
Magnetic field is assumed to be negligible (the strongest measured surface magnetic field strength of a burster is $\lesssim 10^{10}$ G, \citealt{pap11}) in calculations of the opacity.

In order to compute the electron scattering opacity exactly we need to account for induced scattering.
The exact electron scattering opacity can be written as \citep{SPW12} 
\be\label{eq:ESO}
\sigma(x, \mu) = 
\kappa_{\mathrm{e}} \frac{1}{x} \int_0^{\infty} \!\! \!\!  x_1 \mathrm{d}x_1 \int_{-1}^{1}\!\! \mathrm{d} \mu_1 R(x_1,\mu_1;x,\mu)  \left(1+\frac{C I(x_1,\mu_1)}{x_1^3} \right),
\ee
where $R(x,\mu;x_1,\mu_1)$ is the exact angle-dependent relativistic Compton scattering redistribution function (RF; \citealt{AA81,PKB86,NP94a,PS96}) 
that describes the probability for a photon with a dimensionless energy $x$ propagating in a direction defined by $\mu$ 
to be scattered to an energy $x_1$ and to a direction $\mu_1$.
The constant $C$ is defined as
\be
C\!=\!\frac{1}{2 m_{\mathrm{e}}} \left( \frac{h}{m_{\mathrm{e}} c^2} \right)^3.
\ee
Now that $\sigma(x, \mu)$ is formulated, the source function present in the RTE (\ref{eq:RTE}) can be written as a sum of the thermal- and the scattering parts
\begin{align}\begin{split}
S(x,\mu) = &
\frac{k(x)}{\sigma(x,\mu)+k(x)} B_x + \frac{\kappa_{\mathrm{e}}}{\sigma(x,\mu)+k(x)}  \left( 1+ \frac{C I(x,\mu)}{x^3} \right)  \\
& \times x^2 \int_0^{\infty} \frac{\mathrm{d}x_1}{x_1^2} \int_{-1}^{+1} \mathrm{d}\mu_1 R(x,\mu; x_1,\mu_1) I(x_1,\mu_1),
\end{split}\end{align}
where the dimensionless Planck function $B_x$ can be written using the ordinary frequency dependent Planck function $B_{\nu}$ by
\be
B_x = B_{\nu}\frac{\mathrm{d}\nu}{\mathrm{d}x}\!=\!\frac{x^3}{C}\frac{1}{\exp\left(x \frac{m_{\mathrm{e}}c^2}{kT}\right)-1},
\ee
where $k$ is the Boltzmann constant.

The radiation acceleration (see equation~\ref{eq:hydroeq}) is then expressed using the RF as
\be
g_{\mathrm{rad}} = 
\frac{\mathrm{d}P_{\mathrm{rad}}}{\mathrm{d}m} =
\frac{2 \pi}{c} \frac{\mathrm{d}}{\mathrm{d}m} \int_0^{\infty} \mathrm{d}x \int_{-1}^{+1} \mu^2 I(x,\mu) \mathrm{d} \mu
\ee
\be
=\frac{2 \pi}{c} \int_0^{\infty} \mathrm{d}x \int_{-1}^{+1} [\sigma(x,\mu)+k(x)][I(x,\mu)-S(x,\mu)]\mu \mathrm{d}\mu,
\ee
where the derivative with respect to $m$ is replaced by the first moment of the RTE (\ref{eq:RTE}).
The last two main equations are the energy balance equation (that is used to check the solution of RTE)
\be\label{eq:EB}
\int_0^{\infty} \int_{-1}^{+1} [\sigma(x,\mu)+k(x)][I(x,\mu)-S(x,\mu)] \mathrm{d}\mu \!=\! 0,
\ee
and the ideal gas law (see Appendix \ref{sect:appendixA})
\be\label{eq:idealgas}
P_{\mathrm{g}} = N_{\mathrm{tot}}kT,
\ee
where $N_{\mathrm{tot}}$ is the number density of all particles.
In addition, particle and charge conservation are implicitly assumed.

\subsection{Major improvements: level population and opacity calculations}\label{sect:improvements}

Owing to the increasingly important role of metals in the plasma, we have made some major improvements to the code described in \citet{SPW11} and \citet{SPW12} regarding the level population and opacity calculations.
We assume LTE and use the Saha and the Boltzmann equations in order to calculate the number densities of all ionization and excitation states, respectively.
Internal partition functions are now build from the exact energy levels and statistical weights for all ions (up to Ne-like ions with a maximum net charge of $X^{+10}$) and elements up to iron species with a charge of $\Zc \leq 26$ (excluding $\Zc =$ 15, 17, 19 and 21--25).
Previously this was only done for 15 of the most abundant elements\footnote{H, He, C, N, O, Ne, Na, Mg, Al, Si, S, Ar, Ca, Fe and Ni} that we have in the solar composition using only the ground state weights.
In this process of building the exact partition functions, we consider the first 100 atomic excitation energies and statistical weights of states obtained from TOPbase of the Opacity Project\footnote{http://cdsweb.u-strasbg.fr/topbase/home.html}.
Otherwise the internal partition functions are built only from the statistical weights of the ground states like before.
Neglecting the upper excitation levels for some heavier elements ($\Zc > 26$) is not, however, a great source of error due to the rapid truncation of higher order terms in the internal partition function sums via the pressure ionization.
The pressure ionization and level dissolution is accounted for by using the explicit occupation probability formalism \citep{HM88} as described by \cite{Hubeny94} for elements $\Zc \leq 6$ and otherwise the method presented by the Opacity Project \citep{Seaton94}.
Previously the effects from pressure ionization was only taken into account with hydrogen and helium that were the main components of the atmosphere whereas in our new code it is done for every atomic species owing to the high concentration of heavier elements that can affect the general properties of the plasma.
Occupation probability formalism used here is an explicit (ad hoc) method yielding a direct correction factor that can be understood as a combined correction to the statistical weight (to get the real \textit{effective} weight) and to the energy level suppression due to the collective background electric field of the surrounding plasma.

In addition to the exact electron scattering opacity, we take the free-free opacity and the bound-free transitions into account for all elements up to $\Zc = 100$.
Again, it is important to note that when the fraction of metals in the composition increases, all of the opacity sources must be taken into account because of the complicated response they have for the overall opacity picture instead of limiting to only 15 of the most abundant elements.
Especially important are the elements near the iron-peak that have multiple bound-free edges around the spectral peak at energies $\sim$1--10 keV.
The bremsstrahlung opacities of all the ions are calculated assuming that the ion electric field is a Coulomb field of a charge $\Zc e$ equal to the ion charge and using the Gaunt factors from \cite{Su98}.
Bound-free opacities due to photoionization from the ground state of all ions for elements $\Zc \le 30$ is computed using the routine presented in \citet{VY95} and \citet{V96}.
For elements beyond Zn ($\Zc = 30$) we account for the photoionization from the hydrogen-like ions only using the approach by \cite{Karzas61} where the inner nucleus and (non-excited) electrons are replaced by a point-like Coulomb potential.
In addition, photoionization from the first 20 excited levels of all hydrogen-like ions is now included using the same method.
Because of the importance of the bound-free edges to the general properties of the emergent spectrum it is also crucial to consider these excited levels as they, in many cases, smooth the otherwise sharp photoionization edges with additional smaller details below the ionization energy.


\subsection{Method of solution}

The calculations are performed on a logarithmic grid of 100 column depth points $m_{i}$ ranging from $10^{-6}$ to $m_{\mathrm{max}} = 10^5~\mathrm{g}~\mathrm{cm}^{-2}$.
Care is also taken when choosing  $m_{\mathrm{max}}$ to ensure that the effective optical depth $\sqrt{\tau_{\nu,\,\mathrm{b-f},\,\mathrm{f-f}} (m_{\mathrm{max}}) \tau_{\nu}(m_{\mathrm{max}}})$ is larger than unity at all frequencies, 
where $\tau_{\nu,\,\mathrm{b-f},\,\mathrm{f-f}}$ is the optical depth computed with the true opacity only (bound-free and free-free transitions, without scattering).
This condition is necessary for satisfying the inner boundary condition of the radiation transfer problem (diffusion approximation) but it is not trivial as we need to limit our computational grid to the weakly coupled plasma regime (see Appendix \ref{sect:appendixA}).
For the energy grid we use $360$ logarithmically spaced frequency points in the range $10^{14} - 10^{20}~\mathrm{Hz}$ $(\approx4 \times 10^{-4} - 400~\mathrm{keV})$ for the more luminous model atmospheres $(\ell \ge 0.1$), and $10^{14} - 10^{19}~\mathrm{Hz}$ for $\ell < 0.1$.

The course of the calculations is similar to that described in \citet{SPW12}.
First, as an initial guess, a gray atmosphere is constructed for effective temperature $\teff$ (via $\ell$) from pre-tabulated opacity tables.
Then, the radiative acceleration $g_{\mathrm{rad}}$ for all depth points is calculated using a simple and remarkably accurate approximation 
\citep{SPW12}
\be
\frac{g_{\mathrm{rad}}}{g} \approx \ell \left[1+\left(\frac{kT}{40 ~\mathrm{keV}}\right) \right]^{-1}, 
\ee
that accounts for the Klein-Nishina reduction of the electron cross-section.
The radiation pressure $P_{\mathrm{rad}}$ is integrated from $g_{\mathrm{rad}}$ yielding a good starting approximation for the pressure distributions.
Using these starting values, exact opacities are obtained for all depth and frequency points.
Initial starting models built with the aforementioned methods serve as a good starting point for the iterations as the overall flux error is usually already smaller than $30\%$ when compared to the actual converged model.

The formal solution of the radiation transfer equation \eqref{eq:RTE} 
is obtained using the short-characteristic method \citep{OK87} in three angles in each hemisphere.
The full solution is then found via an accelerated $\Lambda$-iteration method (see appendix B in \cite{SPW12} for full description). 
The solution of the radiative transfer equation is then checked against the energy balance equation \eqref{eq:EB} and the surface flux condition
\be
4\pi \int_0^{\infty} H_x(m=0) \mathrm{d}x = \sigma_{\mathrm{SB}}\teff^4,
\ee
where $H_x$ is the local (astrophysical) flux at any given depth point $m$.
Then the temperature corrections for every depth point are computed by using two different error measurements:
the relative error in the flux and the error in the energy balance.
Here the relative flux error is defined
\be
\epsilon_{\mathrm{H}}(m) = 
1 - \frac{H_0}{\int_0^{\infty} H_x(m) \mathrm{d}x},
\ee
where $H_0$ is the correct emergent flux corresponding to a blackbody with a temperature $T_{\mathrm{eff}}$.
In addition, the energy balance error is defined as
\be
\epsilon_{\Lambda}(m) = 
\frac{1}{2} \int_0^{\infty} \mathrm{d}x \int_{-1}^{+1} [\sigma(x, \mu) + k(x)][I(x, \mu) - S(x, \mu)] \mathrm{d}\mu.
\ee
Corrections are then evaluated using a hybrid temperature correction method consisting of three different procedures as follows.
In the deepest layers, the Avrett-Krook flux correction is used based on the relative flux error $\epsilon_{\mathrm{H}}(m)$.
In the intermediate layers, an integral $\Lambda$-iteration method, modified for Compton scattering (based on the energy balance Eq. \eqref{eq:EB}) is used.
This correction is done by finding the necessary temperature change $\delta T_{\Lambda}$ in some particular depth from
\be
\delta T_{\Lambda} = 
-\epsilon_{\Lambda}(m) \left( \int_0^{\infty} \left[ \frac{\Lambda_{\mathrm{d}}(x) - 1}{1- \alpha(x) \Lambda_{\mathrm{d}}(x)} \right] k(x) \frac{\mathrm{d} B_x}{\mathrm{d}T} \mathrm{d}x \right)^{-1},
\ee
where $\Lambda_{\mathrm{d}}(x)$ is the diagonal matrix element of the $\Lambda$-operator and $\alpha(x) = \sigma_{\mathrm{CS}}(x)/(k(x)+\sigma_{\mathrm{CS}}(x))$ and $\sigma_{\mathrm{CS}}$ is the Compton scattering opacity averaged over the relativistic Maxwellian electron distribution (see Eq. (A16) in \cite{PS96} which is equivalent to Eq. \eqref{eq:ESO} if one ignores the induced scattering).
Finally, on the uppermost layers we use the surface correction based on the emerging flux error \citep[see][for a detailed description]{K70}.

As a result of solving these equations iteratively, we obtain a self-consistent NS atmosphere model together with the emergent spectrum of radiation.
This iteration is continued until the relative flux error is less then $0.1\%$ and the relative flux derivative error is smaller then $0.01\%$, in most cases.
For the very luminous models where $g_{\mathrm{rad}}/g \approx 1$ this kind of accuracy is unattainable and the relative flux error can be up to $1-2\%$.

\subsection{Accuracy of computations}

\begin{figure}
\centering
\includegraphics[width=8.5cm]{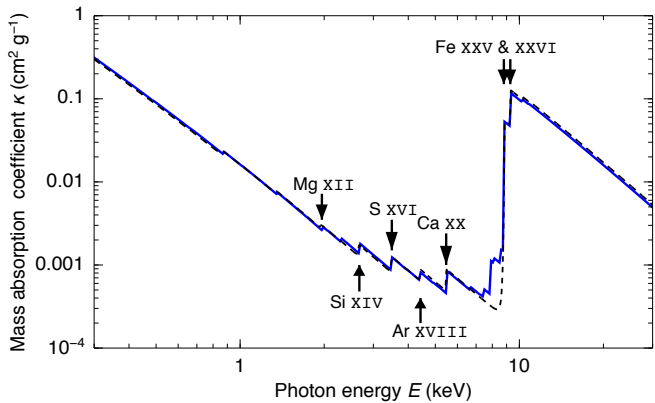}
\caption{\label{fig:opac}
 Comparison of the mass absorption coefficients of the current computations (blue, solid line) and results from \cite{SPW12} (black, dashed line) for solar composition with solar metallicity (SolA1), temperature $T = 10^7$ K and electron pressure $P = 10^{13}$ erg cm$^{-3}$.
  The strongest bound-free edges from different chemical elements are indicated by arrows.
}
\end{figure}

\begin{figure}
\centering
\includegraphics[width=8.5cm]{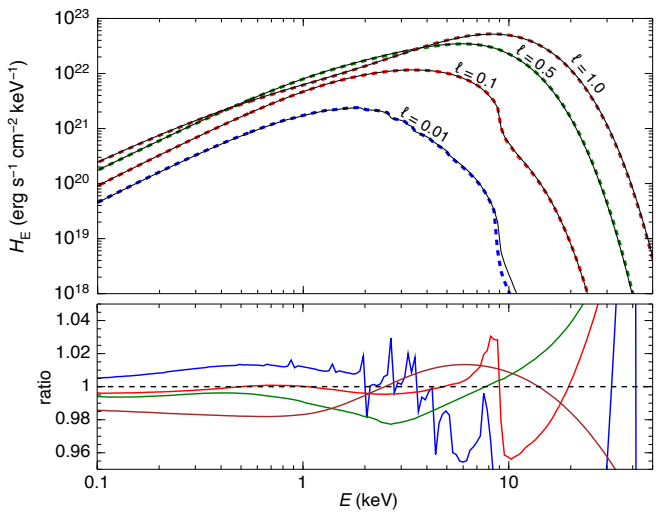}
\caption{\label{fig:sol_comparison}
  Comparison of the emergent spectra for different luminosities of solar composition (SolA1) computed with the current code (dashed, colored lines) and the code presented in \citet{SPW12} (black, solid lines) for $\log g = 14.0$.
  Luminosities shown are $L/L_{\mathrm{Edd}} \equiv \ell = 0.01$ (blue), $0.1$ (red), $0.5$ (green) and $1.0$ (brown).
  The lower panel shows the ratio of these spectra.
}
\end{figure}

Only a few metal-enriched neutron star atmospheres are present in the literature, so comparison and validation of our results is challenging.
We have therefore tested some of our general assumption more thoroughly (see the Appendix \ref{sect:appendixA}), and we have compared our model results of solar composition of elements (SolA1) with the previous work done by \citet{SPW12}.
The differences between the new- and the old models for the opacities obtained for solar composition with $T=10^7~\mathrm{K}$ and $P=10^{13}~\mathrm{erg}\, \mathrm{cm}^{-3}$ (corresponding approximately to the spectral formation depth where $\tau \sim 1$) are presented in Fig.~\ref{fig:opac}.
The overall opacity picture is similar to the previous results by \cite{SPW12}, but as more elements are added the subtle details from additional bound-free edges become visible.
Especially prominent are the additional photoionization edges from the excited states of hydrogen- and helium-like iron below the main ground state edges at $9.278$ keV (Fe \texttt{XXVI}) and $8.828$ keV (Fe \texttt{XXV}).  

These small changes in the opacity profiles then manifest themselves as small differences in the emerging spectra (see Fig.~\ref{fig:sol_comparison}).
The results are, however, in good agreement and largest deviations occur in the very low-luminosity regime where improved level population calculations and additional opacity sources have the strongest effect to the emerging spectra.
When the temperature rises and all elements become fully ionized the deviations due to the different edge strengths slowly vanish and the 
spectra at $\ell = 1.0$ are almost identical.
The small resulting differences are due to slightly altered electron densities.
The discussed differences do not have a strong impact on the previously derived color correction factors because the edge shapes are not modeled anyway by the mere diluted blackbody fit.
However, for models with the increasing abundance of metals that act as sources of true opacity it becomes increasingly important to simulate all the atomic species correctly. 

We also note that no bound-bound opacity is taken into account in the current work.
Because of the partially ionized ions (especially H-, He- and Li-like iron ions) one indeed expects some contribution from the lines to the overall opacity at the lower luminosities ($\ell \lesssim 0.1$ corresponding to about $T_{\mathrm{eff}} \lesssim 20~{\mathrm{MK}}$).
More specifically, the line blanketing would affect the opacity profile the most at the red side of the bound-free edges, rounding the sharp saw-like structures (in addition to the bound-free opacity of the excited states that acts similarly).
In reality, this effect is also coupled with the rotational smearing that acts to smoothen out all of the high-resolution features from the spectra.
Luckily, some computations of atmospheres with lines exists in this temperature range, namely the ones done using the \textit{TMAP} code \citep{RSW08} and the \texttt{carbatm} code \citep[modified version of the code presented in this paper;][]{SKP14}.
In the end, from their results we see that the averaged continuum does not, however, differ much even at $T_{\mathrm{eff}} = 20~\mathrm{MK}$ and some notable contributions from the lines that are strong enough to modify the smoothed spectra are present only at around $10~\mathrm{MK}$ (see e.g. Fig. 8 in \citealt{RSW08}).
We also note that in practice the models for X-ray bursting NS atmospheres are only used up to about $\ell \approx 0.2$ but for completeness we extend our luminosity grid a bit farther down.

\section{New grid of models}\label{sec:newgrid}

\subsection{General properties}\label{sect:gen_props}


\begin{table}\label{tab:comp}
\begin{center}
\caption{Composition of the models with solar ratio of H/He and enhanced metallicities.}
\begin{small}
\begin{tabular}{c c c c c}
  \hline
  Model  &  $X$   &  $Y$  &  $Z$  & $\overline{A}$ \\
  \hline
  SolA1  & 0.738 & 0.249 & 0.013 & 1.26\\
  SolA20 & 0.550 & 0.185 & 0.265 & 1.65\\
  SolA40 & 0.352 & 0.118 & 0.530 & 2.45\\
  \hline  
\end{tabular}
\begin{center}{ 
    Note:
    $X$ is the hydrogen mass fraction,
    $Y$ is the helium mass fraction, 
    $Z$ is the metal mass fraction, and $\overline{A}$ is the mean atomic mass.
}
\end{center}
\end{small}
\end{center}
\end{table}

We have calculated a grid of hot NS atmosphere models using surface gravity values of $\log g = 14.0,~14.3$, and $14.6$, which cover most of the physically realistic NS compactness (see e.g. Fig.~1 in \citealt{SPW11}).
We considered four chemical compositions:
a solar ratio of H/He with metal mass fraction of $Z = 1 Z_{\odot}$ (SolA1) and two enhanced metallicity compositions with $Z = 20 Z_{\odot}$ (SolA20) and $Z = 40 Z_{\odot}$ (SolA40), where the elements from lithium to $\Zc = 100$ are enhanced by factors of 20 and 40, respectively (see Table 1 for specific mass fractions regarding the different models).
The solar abundances were taken from \cite{AGSS09} and for each of the compositions and $\log g$ values we computed models with relative luminosities $\ell$ ranging from $0.001$ to $1.06$ ($\log g = 14.0$), $1.08$ ($\log g = 14.3$) or $1.1$ ($\log g = 14.6$), depending on the chosen surface gravity.
We have also computed a more heuristic composition consisting of pure iron (Fe) only, for comparison.
This kind of atmosphere composition might not be physically realistic, but it serves as an extreme example of a case where the photosphere consists entirely of heavier elements.
It then acts as a proxy that sets the hard lower constraint for $\fc$ for a metal enriched atmosphere.

These metal-enhanced  compositions mimic the conditions in the photosphere of the NS in a simple way when it begins to cool after the rapid nuclear burning of the freshly accreted material.
The key element here is the hydrogen that acts as a catalyst enabling helium burning via the $\alpha$p-process \citep{Schatz11}.
This also explains why we focus on mixed H/He bursts and not pure He where CNO ashes dominate.
Heavy nuclei produced by this $\alpha$p-burning then act as a seed for rp-processes producing even heavier elements up to Te, far exceeding the possible end-results from just $3\alpha$-reaction or CNO-breakout \citep{Schatz11}.
The composition should still be consisting mainly of hydrogen (and helium) but it can now be enriched with heavier nuclear burning ashes if there is some convection in the atmosphere.
Hydrodynamical simulations \citep[see e.g.][]{Woosley04, Fisker08, JMPI10} and (semi-)analytic considerations \citep{WBS06} indeed show that the convective mixing region can approach the photosphere of the NS during the X-ray bursts.
This can then cause an effective mixing of the nuclear burning ashes into the accreted metal-poor mixture of hydrogen and/or helium.

The exact composition of the ashes is not yet well understood and many uncertainties arise from poorly known nuclear reaction rates, which grow into large systematic errors in the end-results of different nucleosynthetic simulations \citep{PJSI13}.
There are general trends indicating that elements around mass number of $\Ac \sim 20 - 30$ (like Si and S) and $\Ac \sim 50 - 60$ (like Fe, Ni and Zn) have considerable overproduction factors when compared to solar abundances \citep[see e.g.][]{KHKF04, Woosley04, Fisker08, JMPI10, Schatz11}.
Because of these uncertainties, we prefer to use the simple metal-enriched models to mimic the presence of the burning ashes, rather than trying to replicate the abundances from the aforementioned works in our atmosphere models.
The solar abundances of metals also peak around medium and heavy elements like carbon, silicon and iron, and therefore the general effects of these metals are likely to resemble the end-results of the rp-process chain  (and that of moderate CNO-burning).
This way we can find the general trends of having metals in the mixture, similar to what was done by \citet{MMJR05} where only iron was considered as an ``average metal''.
Our findings, however, show that it is crucial to the take a broad distribution of these metals into account because they not only contribute to the free-free continuum opacity, but also introduce bound-free edges that can modify the emergent spectrum considerably.
Therefore, our models are likely to give correct general trends of having metals in the photosphere, which yield more electrons and contribute to the total free-free and bound-free opacity.

\begin{figure}
\centering
\includegraphics[width=8.5cm]{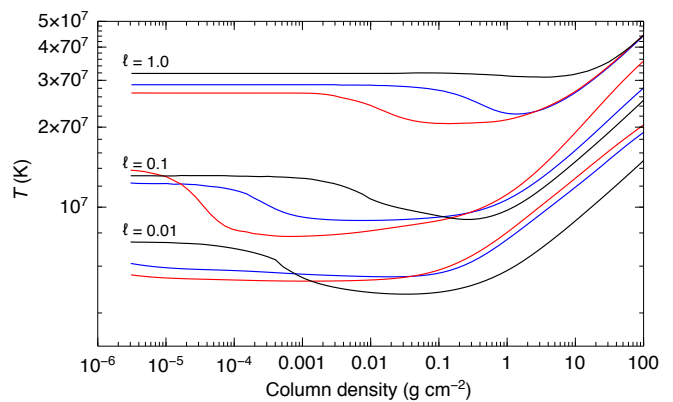}
\caption{\label{fig:example_struc}
Temperature structure for pure iron (red), solar ratio of H/He with $1$ (black) and $40$ (blue) times the solar metallicity $Z_{\odot}$ for luminosities $l = 0.01$, $0.1$ 
and $1.0$ for $\log g = 14.0$. 
}
\end{figure}

\begin{figure}
\centering
\includegraphics[width=8.5cm]{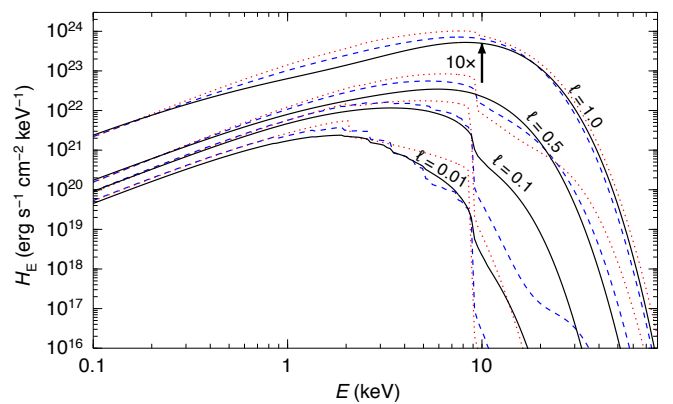}
\caption{\label{fig:example_spec}
  Emergent spectra for pure iron (red, dotted line), solar ratio of H/He with metallicity $Z = 1 Z_{\odot}$ (black solid lines) and $Z =40 Z_{\odot}$ (blue dashed line) for luminosities $\ell = 0.01$, $0.1$ $0.5$ and $1.0$ computed with $\log g = 14.0$.
  For clarity models corresponding to luminosity of $\ell = 1.0$ are shifted along the ordinate axis by a factor of $10$.
}
\end{figure}

Examples of the temperature structures and emergent spectra for models with $\log g = 14.0$ and the selected chemical compositions are shown in Figs.~\ref{fig:example_struc} and \ref{fig:example_spec}.
From the temperature structure of the models (see Fig.~\ref{fig:example_struc}) one can see the general trend that more metal-rich atmospheres have cooler outer 
layers.\footnote{For the same $\ell$ the metal-rich models have higher $T_{\mathrm{eff}}$ because of the opacity dependence on $X$ (see equations \ref{eq:kappae} and \ref{eq:fedd}).
This implies that the effect is actually even stronger than that seen in Fig.~\ref{fig:example_struc}.}
The temperature in the upper optically thin atmospheric layers is determined by the balance between heating and cooling of the matter by radiation. 
The absorption opacity in the most parts of the upper layers is insignificant because of the low density, and the temperature is equal to the Compton temperature determined by heating and cooling by non-coherent electron scattering only (see detail discussions in \citealt{LST86,PSZ91}). 
In fact, the surface temperature is close to the color temperature of the emergent spectrum, $T_{\rm c} = f_{\rm c} T_{\rm eff}$. 
The temperature of the deeper layers will be mainly determined by the balance between photon absorption $\int_0^\infty k_\nu J_\nu\, d\nu$ and emission $\int_0^\infty k_\nu B_\nu\, d\nu$, if the relative luminosity $\ell$ is not too high. The absorption opacity increases toward lower photon energies and the balance  occurs at temperatures which are much smaller than the color temperature of the emergent spectrum.  
This leads to a temperature minimum at some depths where the  cooling and heating are dominated by true absorption and corresponding thermal emission. 
The depth and the width of this temperature depression depend on the ratio between the absorption and the scattering opacities in the atmosphere.
Absorption is stronger in atmospheres with more abundant heavy elements and higher gravity (which results in a  higher density). 
Indeed, the largest width of the temperature depression is seen for pure iron atmospheres. 
In the optically thick layers, the temperature increases according to the diffusion approximation $T^4 \propto \tau_{\rm R}$, where $\tau_{\rm R}$ is the Rosseland optical depth.  

The emerging spectra are almost always harder than the corresponding blackbody of temperature $T_{\mathrm{eff}}$ (see Fig~\ref{fig:example_spec}).
Only in the case of the largest surface gravity and pure iron composition can the bound-free edges suppress the radiation enough, so that the emerging radiation becomes softer.
At very high luminosities $\ell \gtrsim 0.8$, we confirm the previous results that the spectra become harder because of the increasing contribution from the radiation acceleration when $g_{\mathrm{rad}}/g \approx 1$ \citep{London86, LST86, Ebi87, PSZ91}.
It leads to a decrease in the matter density and the absorption-to-scattering opacity ratio. 
Therefore, the emergent photons are born at larger depths and at larger temperatures.
At lower relative luminosity $\ell$ (i.e. lower effective temperature), the density and the absorption opacity increases.
This leads to  the appearance of partially ionized species, first of all H- and He-like iron ions.   
At these lower relative luminosities, the absorption edges of the H-like Fe\texttt{XXVI} (9.278 keV) and He-like Fe\texttt{XXV} (8.828 keV) ions strongly 
affect the emerging radiation.
At the coldest temperatures (around $\ell \sim 0.01 - 0.1$) this picture is complicated even further by numerous additional bound-free edges from various other chemical elements present in the mixture.

\subsection{Dilution and color correction factors}\label{sec:applic}

The model spectra are very close to a blackbody shape in the observed \textit{RXTE}/PCA \citep{JMR06} energy band and it is very common that the actual observations (especially of X-ray bursts) are also fitted with a (diluted) blackbody function \citep[see e.g.][]{GMH08}.
The diluted blackbody has two parameters: the color temperature and the normalization 
\be\label{eq:Kfc}
   K \equiv \left( \frac{R_{\rm bb}\,{\rm (km)}}{D_{10}} \right) ^2 = \frac{1}{f_{\mathrm{c}}^4} \left( \frac{R\,{\rm (km)}\, (1+z) }{D_{10}}\right) ^2 ,
\ee
where $R_{\rm bb}$ is the blackbody radius, $D_{10}$ is the distance to the source in units of 10 kpc and $z$ is gravitational redshift at the NS surface. 
The theoretical evolution of $f_{\mathrm{c}}$ can be related to the observed evolution of $K^{-1/4}$ with flux during the cooling tail \citep{Penninx89,vP90,SPRW11,SPW11} and the constraints on the NS mass and radius can be thus obtained.

\begin{figure}
\centering
\includegraphics[width=8.5cm]{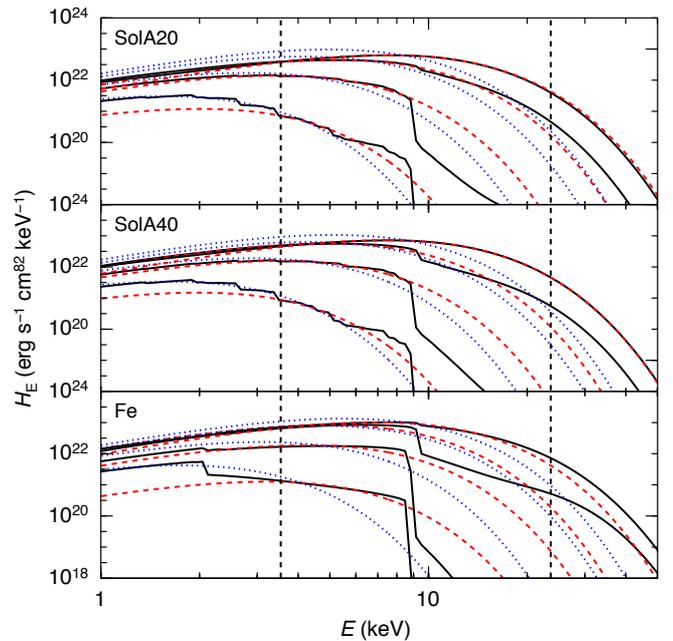}
\caption{\label{fig:dbb}
  Emergent spectra (black, solid line) for different relative luminosities of $\ell = 0.01$, $0.1$, $0.5$ and $1.0$ for compositions of the solar ratio of H/He with $Z = 20Z_{\odot}$ (upper panel) and $Z = 40Z_{\odot}$ (middle panel); and pure iron (bottom panel) for $\log g = 14.0$ is shown.
  Best-fit diluted blackbody function (red, dashed line) is also shown together with a blackbody function where $T_{\mathrm{bb}} = T_{\mathrm{eff}}$ (blue, dotted line).
  Vertical dashed lines correspond to the observed $(3-20) \times (1+z)$ keV range where the fitting is performed.
  }
\end{figure}

\begin{figure}
\centering
\includegraphics[width=8.5cm]{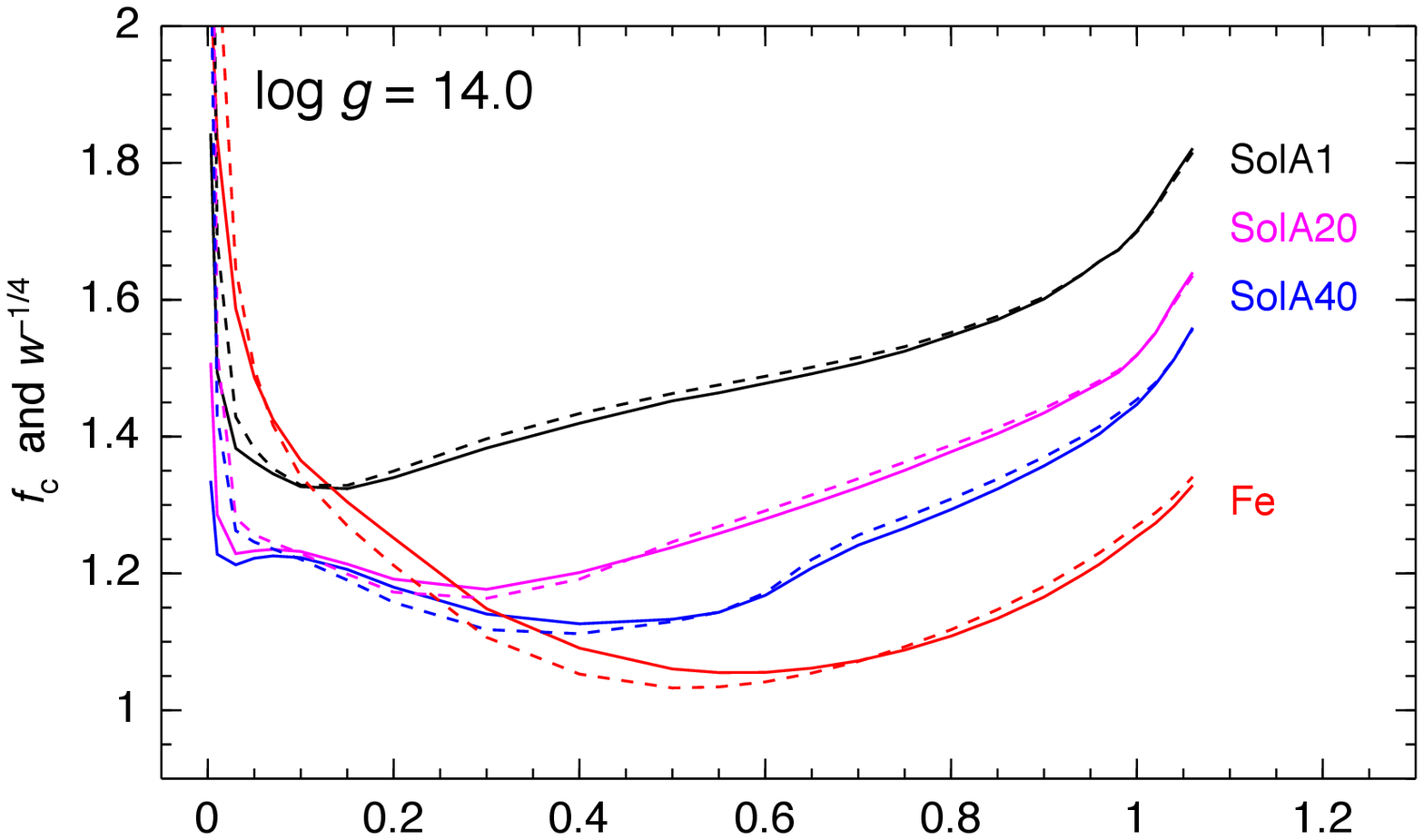}
\includegraphics[width=8.5cm]{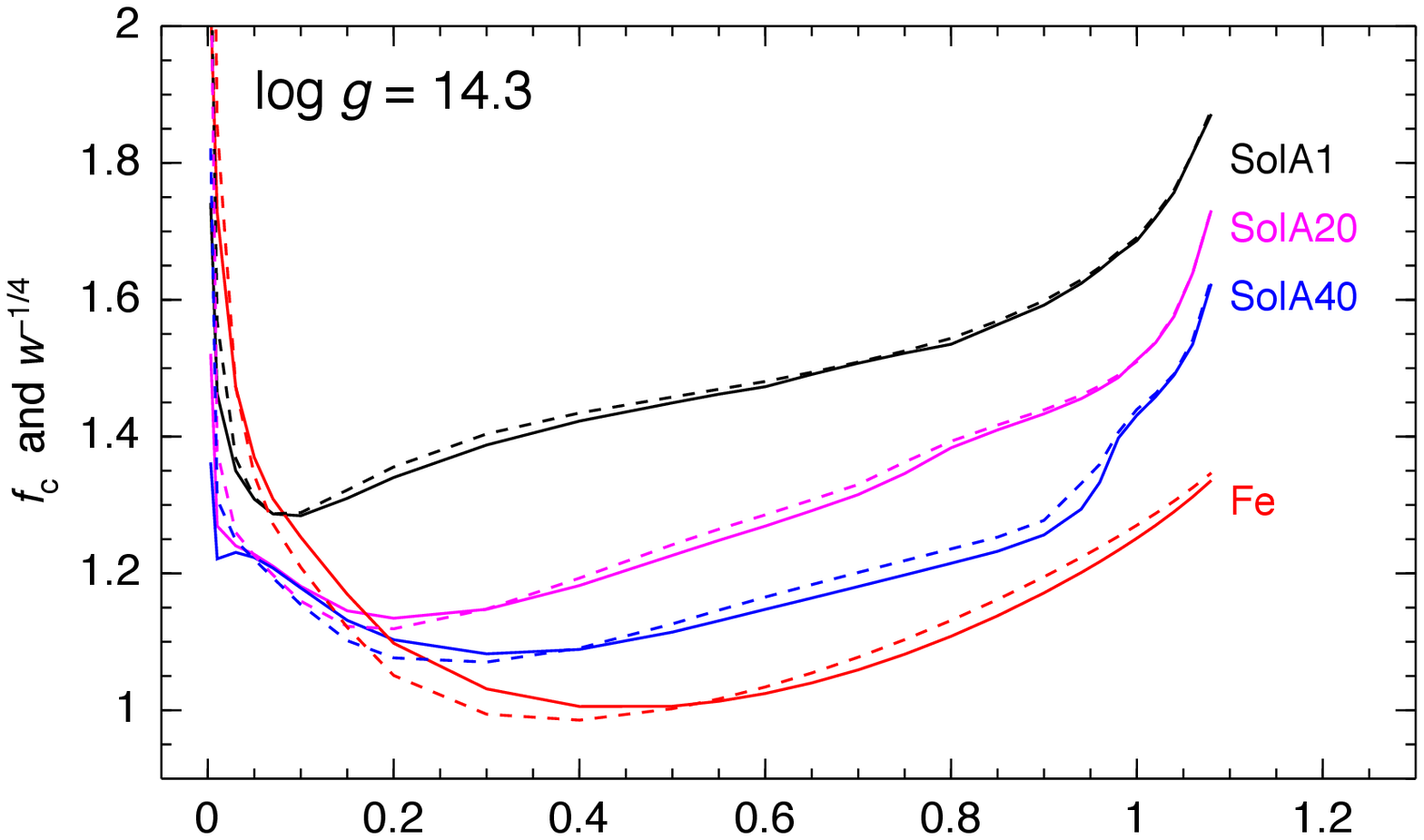}
\includegraphics[width=8.5cm]{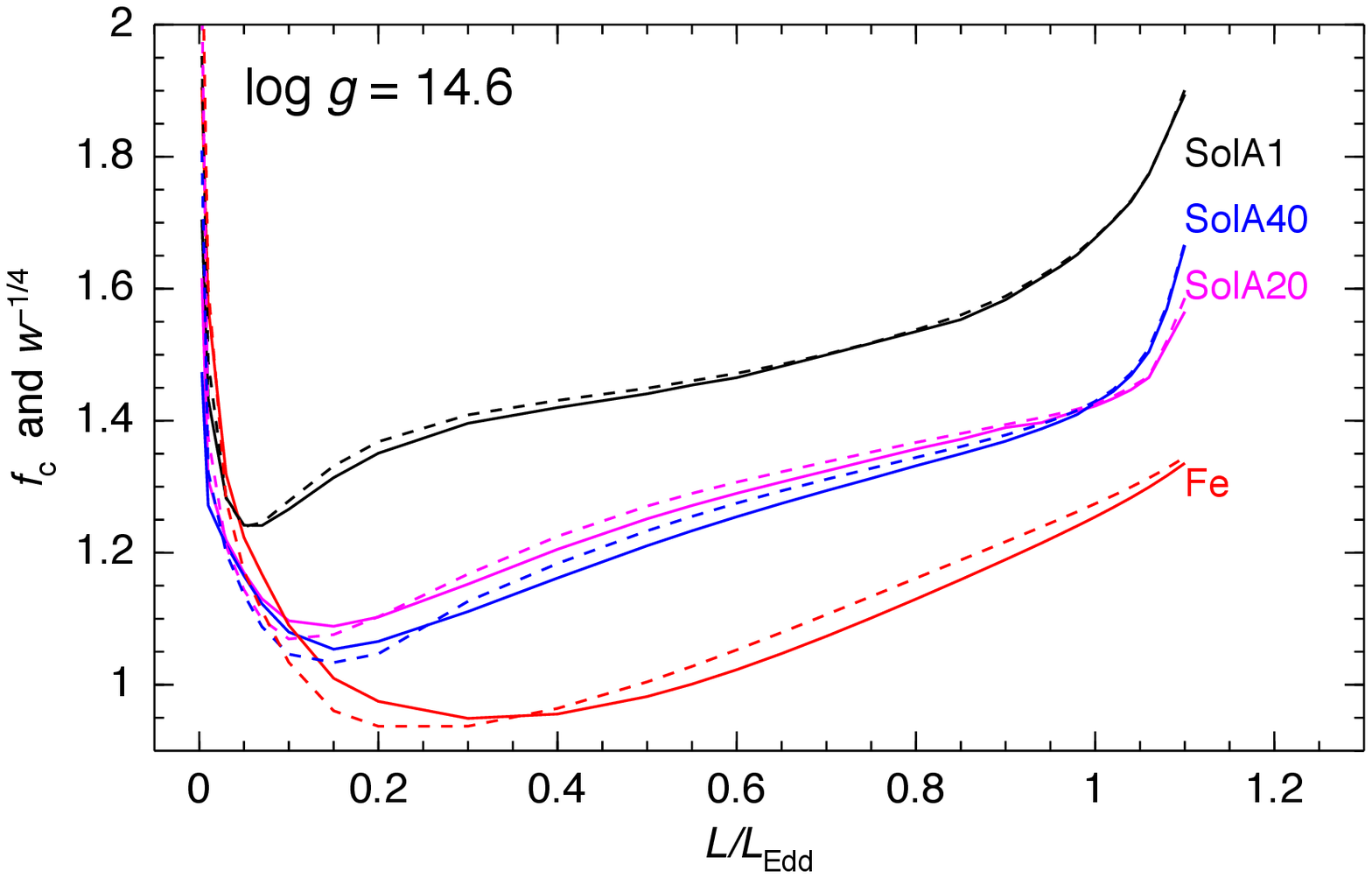}
\caption{\label{fig:w-l}
  Color correction factors $f_{\mathrm{c}}$ (solid lines) and dilution factors $w^{-1/4}$ (dashed lines) for model atmospheres consisting of pure iron (red) and solar mixture of H/He with $Z = Z_{\odot}$ (SolA1, black), $Z = 20Z_{\odot}$ (SolA20, magenta) and $Z = 40Z_{\odot}$ (SolA40, blue) against the relative luminosity $\ell$ for $\log g = 14.0$ (upper panel), $\log g = 14.3$ (middle panel) and $\log g = 14.6$ (bottom panel).
}
\end{figure}

In order to make our models easily accessible for data analysis we computed the so-called color correction curves \citep[see e.g.][]{SPW11,SPW12} 
for our new set of models.
These can then be directly compared with observed best-fit values for the blackbody normalization if the emission is coming from a passively cooling NS surface \citep[see e.g.][]{SPW12,PNK14, KNL14}.
All theoretical emergent spectra were fitted with a diluted blackbody 
\be
F_E \approx w B_E (f_{\mathrm{c}} T_{\mathrm{eff}}),
\ee
where $f_{\mathrm{c}}$ is the color correction (or hardness) factor and $w$ is the dilution factor.
Because of the normalization of the diluted blackbody function we have an approximate relation $\fc \approx w^{-1/4}$ between these two variables.\footnote{$\int_0^{\infty} w B_E(f_{\mathrm{c}} T) ~dE \propto w f_{\mathrm{c}}^{4} B(T)$ from where we require $w f_{\mathrm{c}}^4 = 1$.}
The fitting was done using the first method described in \cite{SPW11} in the energy band $(3-20) \times (1+z)~\mathrm{keV}$ corresponding to the observed \textit{RXTE}/PCA detector energy band (see Fig. \ref{fig:dbb}).
Value of the redshift is calculated from the $\log g$ values by adopting NS mass of $1.4~M_{\odot}$ corresponding to values of $z=0.18$, $0.27$ and $0.42$
for $\log g=14.0$ ($R=14.80~\mathrm{km}$), $14.3$  ($R=10.88~\mathrm{km}$) and $14.6$  ($R=8.16~\mathrm{km}$), respectively.
Evolution of the color correction and dilution factors are illustrated in Fig. \ref{fig:w-l} and the results of the fitting are presented in Table \ref{tab:fc-w-l} for SolA20 ($\log g = 14.0$) as an example.
Results for other chemical compositions and gravities are given in the Table \ref{tab:fc-w-l_A}.

\begin{table}
\caption{Color correction and dilution factors from the blackbody fits to the spectra of SolA20 atmosphere models at $\log g=14.0$.}
\centering
\begin{tabular}[c]{c c c c c c}
  \multicolumn{6}{c}{$\quad$ $Z=20Z_{\odot}$ $\quad$ $\log g = 14.0$ $\quad$ $T_{\mathrm{Edd}}= 1.75$ keV} \\
  \multicolumn{6}{c}{$R=14.80$ km $\quad$ $z=0.18$} \\
\hline
\hline
$\ell$ & $g_{\mathrm{rad}}/g$ & $T_\mathrm{eff}$ (keV) & $f_{\mathrm{c}}$ & $w$ & $w f_{\mathrm{c}}^{4}$ \\
\hline
0.001 & 0.017 & 0.311 & 1.805 & 0.002 & 0.023 \\
0.003 & 0.026 & 0.409 & 1.508 & 0.023 & 0.121 \\
0.010 & 0.043 & 0.553 & 1.286 & 0.185 & 0.505 \\
0.030 & 0.067 & 0.728 & 1.229 & 0.371 & 0.847 \\
0.050 & 0.087 & 0.827 & 1.233 & 0.401 & 0.927 \\
0.070 & 0.106 & 0.900 & 1.235 & 0.416 & 0.968 \\
0.100 & 0.139 & 0.984 & 1.232 & 0.438 & 1.008 \\
0.150 & 0.190 & 1.089 & 1.214 & 0.484 & 1.049 \\
0.200 & 0.240 & 1.170 & 1.192 & 0.529 & 1.067 \\
0.300 & 0.332 & 1.295 & 1.177 & 0.546 & 1.046 \\
0.400 & 0.414 & 1.391 & 1.201 & 0.496 & 1.033 \\
0.500 & 0.526 & 1.471 & 1.238 & 0.415 & 0.975 \\
0.550 & 0.568 & 1.507 & 1.258 & 0.386 & 0.968 \\
0.600 & 0.610 & 1.540 & 1.280 & 0.359 & 0.964 \\
0.650 & 0.651 & 1.571 & 1.302 & 0.335 & 0.962 \\
0.700 & 0.692 & 1.600 & 1.326 & 0.312 & 0.962 \\
0.750 & 0.734 & 1.628 & 1.351 & 0.290 & 0.966 \\
0.800 & 0.775 & 1.655 & 1.378 & 0.270 & 0.972 \\
0.850 & 0.816 & 1.680 & 1.404 & 0.251 & 0.975 \\
0.900 & 0.857 & 1.704 & 1.435 & 0.232 & 0.982 \\
0.940 & 0.890 & 1.723 & 1.464 & 0.216 & 0.990 \\
0.960 & 0.907 & 1.732 & 1.478 & 0.208 & 0.991 \\
0.980 & 0.921 & 1.741 & 1.494 & 0.199 & 0.993 \\
1.000 & 0.939 & 1.750 & 1.519 & 0.188 & 1.000 \\
1.020 & 0.953 & 1.758 & 1.551 & 0.173 & 0.998 \\
1.040 & 0.972 & 1.767 & 1.599 & 0.154 & 1.008 \\
1.060 & 0.989 & 1.775 & 1.634 & 0.140 & 1.012 \\
\hline
\end{tabular}
\label{tab:fc-w-l}
\begin{center}
{ 
  \textbf{Notes.} Results for other chemical compositions and surface gravities are listed in the Appendix \ref{sect:appB}.
}
\end{center}
\end{table}

A common feature of the color correction factors is their non-monotonic dependence on the relative luminosity $\ell$.
The color correction factors have a maxima at the maximum possible $\ell$, after which the correction starts to decrease as the luminosity goes down.
As the relative luminosity decreases further, a local minima is finally reached around $\ell = 0.1 - 0.6$.
The position of the minimum and its strength strongly depends on the metallicity. 
The  minimum shifts toward larger $\ell$, when the heavy element abundance increases:
for the solar abundance the minimum occurs at $\ell \approx 0.1$ whereas for the pure iron atmospheres it is located at $\ell \approx0.6$. 
This happens because the iron edges near 9 keV suppress the high-energy radiation, forcing the most effective cooling window into lower energies and thus 
reducing the color correction factor.
The efficiency of this process depends on the occupation number of the H- and He-like iron ions.
These occupation numbers, on the other hand,  are determined by the iron abundances and the ratio of the total number of iron atoms in different stages of ionization, which, in turn, is depended on the temperature and the matter density.
It means that similar absorption edges in the emergent spectra can arise at low temperature and low iron abundance as well as at a relatively high
temperature and an increased iron abundance.
This causes the shift of the local $f_{\rm c}$ minimum to larger $\ell$, when the iron abundance is increased.  
It also explains the same track of the color corrections at the low luminosities for the metallicities of $Z = 20Z_{\odot}$ and $40 Z_{\odot}$ because once the edges are strong enough to suppress the high-energy radiation, the spectra appear rather similar.
It is only when the peak of the effective Planck function $B_{{E}}(T_{\mathrm{eff}})$ can climb over this edge that the color correction starts to increase again with increasing $\ell$.
In the very low-luminosity regime ($\ell < 0.1$) the emerging spectra are hardened by the free-free absorption as first noted by \cite{London84, London86}.
At these low photon energies the free-free opacity dominates over the electron scattering and causes additional hardening due to the strong $\propto E^{-3}$ energy dependency.
The effect is then further enhanced by the decreasing effective temperature $T_{\mathrm{eff}}$ leading to an even harder spectrum.


Another notable feature is the apparent super-Eddington luminosities (when the relative luminosity is $\ell > 1$) as first noted by \cite{SPW12}.
This happens because of the Klein-Nishina reduction of the electron cross-sections.
Because we compute our models all the way up to $g_{\mathrm{rad}}/g \approx 1$, formally super-Eddington luminosities are needed to counter the fact that the Eddington luminosity is defined for the Thomson cross-section.
As Klein-Nishina reduction depends on the electron temperature it also makes the upper limiting luminosity dependent on $\log g$ as the models with stronger surface gravity tend to be hotter.
We also note that in our definition of the Eddington luminosity (and hence also in the Eddington temperatures tabulated) we use the approximation $\ke \approx 0.2(1+X)~\mathrm{cm}^2~\mathrm{g}^{-1}$ for the Thomson scattering opacity.
This approximation assumes full ionization and that the ratio of number of electrons to protons and neutrons in metals is half, which, in reality, is not the case.
In practice, these corrections are, however, rather small and easy to apply, so for clarity and convenience we use the standard definition of the Eddington luminosity defined by equations \eqref{eq:ledd} and \eqref{eq:kappae}.

The new models also show significant dependency on the surface gravity because of the metals.
When increasing the $\log g$ value the density of the photosphere is also enhanced.
This increase in density, in turn, leads to smaller iron ionization fractions and, therefore, to larger bound-free edges as the number density of absorbing elements increase.
The luminosity where the large iron edges are finally seen to vanish (due to almost full ionization) can be traced to $\ell=0.5$ for SolA20 and $\ell=0.6$ for SolA40 with $\log g = 14.0$ whereas the same limiting luminosity appear around $\ell=0.8$ for SolA20 and $\ell=0.9$ for SolA40 when the surface gravity is increased to $\log g = 14.3$.
The disappearance of these edges then lead to a break in the color correction curves when photons are able to escape from the blue side of the photoionization edge.
This change results in a sudden increase in the hardness of the emerging spectra and hence increase in the $f_{\mathrm{c}}$.
For the $\log g = 14.6$ this break is shifted to near the Eddington luminosity and coincides with the color correction increasing when the luminosity becomes close to the Eddington limit. 
These two effects in turn make the color correction evolution for models with $\log g = 14.6$ very sharp at $\ell > 0.8$.
Moreover, the color correction evolutions $\ell - f_{\mathrm{c}}$ for both metal-enriched compositions are very similar when $\log g =14.6$.
The similarity originates because the increased temperature with SolA40 model (that will reduce the number of absorbing nuclei) is countered by the larger initial number of iron nuclei.
As a result the model spectra of SolA20 and SolA40 became self-similar as the SolA20 model, in contrast, has a lower temperature (i.e. more non-ionized nuclei that can absorb) but fewer heavy element nuclei (like iron) capable of participating in the absorption processes.
The $\ell - f_{\mathrm{c}}$ curves for pure iron models do not show any significant increase because iron is not fully ionized even at $\ell \sim 1$.

From our color correction fits we can limit the variability of $f_{\mathrm{c}}$ to be between about $1.3 - 1.8$ near the Eddington limit and $1.0 - 1.4$ around $\ell \approx 0.5$.
At lower luminosities of around $\ell \approx 0.1$ the scatter is even smaller ($\fc \approx 1.2 - 1.4$) but there is a possibility that at this late stage the spectra of the cooling NS surface are already likely to be contaminated by additional heating from the in-falling gas from the accretion flow. 
It should also be noted that the lower limit set by the pure iron composition is most likely not realistic as at least some hydrogen (and other elements too from the ashes of the previous bursts) is likely to be present in the atmosphere.
Still, it acts as a heuristic limiting case, with a maximum concentration of metals.
It is also interesting to note that even small amounts of metals are able to modify the color correction factor substantially. 

It is also clear that the spectra of NS with pure iron envelopes and even just strongly enhanced heavy element abundances cannot be fitted well with smooth blackbody functions.
Therefore, our computations of color factors for their model spectra are rather arbitrary.
In fact, when comparing to observations we would expect to find significant residuals near iron absorption edges if the atmosphere indeed has large (heavy) metal content.
However, we have to also keep in mind that various reasons, such as the fast rotation of the NS, additional spectral smearing because of the rapidly rotating (spreading) boundary layer or additional radiation from the optically thin recombination continua of iron ions arising in a surrounding envelope can reduce the observed strength of the iron absorption edges.

In addition to the direct observations of the iron edges, the connection between the color correction factor and the blackbody normalization (given by equation~\ref{eq:Kfc}) opens up a possibility of observing the presence of these metals indirectly:
even a small increase in their abundance would end up decreasing the color correction factor toward unity.
Our results show that this mechanism is enough to explain the small burst-to-burst variations observed in the cooling tracks between separate bursts from the same source \citep[see e.g.][]{GPO12, PNK14}. On the other hand, abnormally large deviations from the typical cooling track for some individual burst might indicate an especially large ash content and/or effective mixing in the photosphere.
In addition, these outlier bursts are expected to show strongest residuals (i.e. poor $\chi^2$ values) when the photosphere cools giving yet another testable prediction of the models.
These connections to the observations remain to be tested and will be pursued in future publications.

\section{Summary}\label{sec:summary}

We have presented a new NS atmosphere modeling code which is an extension of our previous works \citep{SPW11, SPW12}. 
The code is especially designed to model the emerging spectra from X-ray bursting NSs with metal-rich atmospheres and can be used for NS mass and radius estimates.
The computational scheme and the code are validated producing results in a good agreement with our previous work.

We have assumed an ideal plasma in LTE.
The model atmospheres are considered to be in hydrostatic and radiative equilibrium. 
We use the standard plane-parallel atmosphere approximation. 
No magnetic field nor the stellar rotation are taken into account.
For Compton scattering, the exact angle-dependent relativistic redistribution kernel is used.
As a new addition, the bound-free and free-free opacities are now taken into account for almost all (up to $\Zc = 100$) elements. 
We also use exactly constructed internal partition functions for calculations of the ionization fractions and use the occupation formalism to take the pressure ionization effects into account in comparison to previous work.

To study the effects arising from the metals we have computed a new set of hot NS atmospheres with enhanced metallicities.
Atmospheric structures and emergent spectra are calculated for four different chemical compositions (pure iron and solar hydrogen/helium mix with various abundances of heavy elements of $Z = 1$, $20$ and $40\, Z_{\odot}$) and for three different surface gravities of $\log g = 14.0$, $14.3$ and $14.6$.
The relative luminosities range from $\ell = 0.001$ to $1.06 - 1.10$ (depending on $\log g$).

All computed emergent spectra differ from the blackbody spectra.
The spectral shape for the relatively luminous models is, however, well reproduced by a diluted blackbody  that is used to fit all of our models in the observed \textit{RXTE}/PCA energy band ($3-20$ keV).%
\footnote{The obtained color correction and dilution factors can also be found at the CDS.}
The spectra for the less luminous models have significant iron absorption edges, and their approximation by a diluted blackbody becomes worse.

We found that the emergent spectra are strongly dependent on the fraction of (heavy) metals in the composition. 
We constrain the color correction factor to be between $f_{\mathrm{c}} \approx 1.3 - 1.8$ near $\ell \approx 1$ and $\fc \approx 1.0 - 1.4$ around $\ell \approx 0.5$, depending on the metallicity.
These limits show that by varying the fraction of the burning ashes in the composition, the color correction can change quite considerably.
It also shows that -- in addition to the uncertain detection of the photoionization edges in the emergent spectra -- the color correction evolution can be used as a new tool to probe the nuclear burning (and the mixing via convection) at the NS surface due to the strong metal dependency.

\begin{acknowledgements}
Authors would like to thank Erik Kuulkers and Jan-Uwe Ness for valuable discussions.
This research was supported by the V\"ais\"al\"a Foundation and by the University of Turku Graduate School in Physical and Chemical Sciences (JN).
VS was supported by the German Research Foundation (DFG) grant WE 1312/48-1.
JJEK acknowledges support from the ESA research fellowship programme.
JP acknowledges funding from the Academy of Finland grant 268740.
JN and JJEK acknowledge support from the Faculty of the European Space Astronomy Centre (ESAC).
We thank the International Space Science Institute (ISSI) located in Bern, Switzerland, 
for sponsoring an International Team on type I X-ray bursts
where this project has started.
This research was undertaken on Finnish Grid Infrastructure (FGI) resources.
\end{acknowledgements}

\bibliographystyle{aa}

\begin{thebibliography}{62}
\expandafter\ifx\csname natexlab\endcsname\relax\def\natexlab#1{#1}\fi

\bibitem[{{Aharonian} \& {Atoyan}(1981)}]{AA81}
{Aharonian}, F.~A. \& {Atoyan}, A.~M. 1981, \apss, 79, 321

\bibitem[{{Asplund} {et~al.}(2009){Asplund}, {Grevesse}, {Sauval}, \&
  {Scott}}]{AGSS09}
{Asplund}, M., {Grevesse}, N., {Sauval}, A.~J., \& {Scott}, P. 2009, \araa, 47,
  481

\bibitem[{{Bezanson} {et~al.}(2012){Bezanson}, {Karpinski}, {Shah}, \&
  {Edelman}}]{julia2012}
{Bezanson}, J., {Karpinski}, S., {Shah}, V.~B., \& {Edelman}, A. 2012, 
  \texttt{[ArXiv:1209.5145]}

\bibitem[{{Bhattacharyya} {et~al.}(2010){Bhattacharyya}, {Miller}, \&
  {Galloway}}]{BMG10}
{Bhattacharyya}, S., {Miller}, M.~C., \& {Galloway}, D.~K. 2010, \mnras, 401, 2

\bibitem[{{Chen}(1984)}]{Chen84}
{Chen}, F.~F. 1984, Introduction to Plasma Physics and Controlled Fusion
  (Heidelberg: Springer)

\bibitem[{{Cumming} \& {Bildsten}(2000)}]{CB00}
{Cumming}, A. \& {Bildsten}, L. 2000, \apj, 544, 453

\bibitem[{{Ebisuzaki}(1987)}]{Ebi87}
{Ebisuzaki}, T. 1987, \pasj, 39, 287

\bibitem[{{Fisker} {et~al.}(2008){Fisker}, {Schatz}, \&
  {Thielemann}}]{Fisker08}
{Fisker}, J.~L., {Schatz}, H., \& {Thielemann}, F.-K. 2008, \apjs, 174, 261

\bibitem[{{Fujimoto} {et~al.}(1981){Fujimoto}, {Hanawa}, \& {Miyaji}}]{FHM81}
{Fujimoto}, M.~Y., {Hanawa}, T., \& {Miyaji}, S. 1981, \apj, 247, 267

\bibitem[{{Fushiki} \& {Lamb}(1987)}]{FL87}
{Fushiki}, I. \& {Lamb}, D.~Q. 1987, \apjl, 323, L55

\bibitem[{{Galloway} {et~al.}(2008){Galloway}, {Muno}, {Hartman}, {Psaltis}, \&
  {Chakrabarty}}]{GMH08}
{Galloway}, D.~K., {Muno}, M.~P., {Hartman}, J.~M., {Psaltis}, D., \&
  {Chakrabarty}, D. 2008, \apjs, 179, 360

\bibitem[{{G{\"u}ver} {et~al.}(2012){G{\"u}ver}, {Psaltis}, \&
  {{\"O}zel}}]{GPO12}
{G{\"u}ver}, T., {Psaltis}, D., \& {{\"O}zel}, F. 2012, \apj, 747, 76

\bibitem[{{Haensel} {et~al.}(2007){Haensel}, {Potekhin}, \& {Yakovlev}}]{HPY07}
{Haensel}, P., {Potekhin}, A.~Y., \& {Yakovlev}, D.~G. 2007, Astrophysics and
  Space Science Library, Vol. 326, {Neutron Stars 1: Equation of State and
  Structure} (New York: Springer)

\bibitem[{{Hoffman} {et~al.}(1980){Hoffman}, {Cominsky}, \& {Lewin}}]{HCL80}
{Hoffman}, J.~A., {Cominsky}, L., \& {Lewin}, W.~H.~G. 1980, \apjl, 240, L27

\bibitem[{{Hubeny} {et~al.}(1994){Hubeny}, {Hummer}, \& {Lanz}}]{Hubeny94}
{Hubeny}, I., {Hummer}, D.~G., \& {Lanz}, T. 1994, \aap, 282, 151

\bibitem[{{Hummer} \& {Mihalas}(1988)}]{HM88}
{Hummer}, D.~G. \& {Mihalas}, D. 1988, \apj, 331, 794

\bibitem[{{Ibragimov} {et~al.}(2003){Ibragimov}, {Suleimanov}, {Vikhlinin}, \&
  {Sakhibullin}}]{I03}
{Ibragimov}, A.~A., {Suleimanov}, V.~F., {Vikhlinin}, A., \& {Sakhibullin},
  N.~A. 2003, Astronomy Reports, 47, 186

\bibitem[{{in't Zand} \& {Weinberg}(2010)}]{iZW10}
{in't Zand}, J.~J.~M. \& {Weinberg}, N.~N. 2010, \aap, 520, A81

\bibitem[{{Jahoda} {et~al.}(2006){Jahoda}, {Markwardt}, {Radeva}, {Rots},
  {Stark}, {Swank}, {Strohmayer}, \& {Zhang}}]{JMR06}
{Jahoda}, K., {Markwardt}, C.~B., {Radeva}, Y., {et~al.} 2006, \apjs, 163, 401

\bibitem[{{Jos{\'e}} {et~al.}(2010){Jos{\'e}}, {Moreno}, {Parikh}, \&
  {Iliadis}}]{JMPI10}
{Jos{\'e}}, J., {Moreno}, F., {Parikh}, A., \& {Iliadis}, C. 2010, \apjs, 189,
  204

\bibitem[{{Joss}(1978)}]{Joss78}
{Joss}, P.~C. 1978, \apjl, 225, L123

\bibitem[{{Kajava} {et~al.}(2014){Kajava}, {N{\"a}ttil{\"a}}, {Latvala},
  {Pursiainen}, {Poutanen}, {Suleimanov}, {Revnivtsev}, {Kuulkers}, \&
  {Galloway}}]{KNL14}
{Kajava}, J.~J.~E., {N{\"a}ttil{\"a}}, J., {Latvala}, O.-M., {et~al.} 2014,
  \mnras, 445, 4218

\bibitem[{{Karzas} \& {Latter}(1961)}]{Karzas61}
{Karzas}, W.~J. \& {Latter}, R. 1961, \apjs, 6, 167

\bibitem[{{Koike} {et~al.}(2004){Koike}, {Hashimoto}, {Kuromizu}, \&
  {Fujimoto}}]{KHKF04}
{Koike}, O., {Hashimoto}, M.-a., {Kuromizu}, R., \& {Fujimoto}, S.-i. 2004,
  \apj, 603, 242

\bibitem[{{Kurucz}(1993)}]{K93}
{Kurucz}, R. 1993, CD-ROMs Cambridge, Mass.: Smithsonian Astrophysical
  Observatory, 11

\bibitem[{{Kurucz}(1970)}]{K70}
{Kurucz}, R.~L. 1970, SAO Special Report, 309

\bibitem[{{Lapidus} {et~al.}(1986){Lapidus}, {Syunyaev}, \&
  {Titarchuk}}]{LST86}
{Lapidus}, I.~I., {Syunyaev}, R.~A., \& {Titarchuk}, L.~G. 1986, Soviet
  Astronomy Letters, 12, 383

\bibitem[{{Lattimer} \& {Prakash}(2004)}]{LP04}
{Lattimer}, J.~M. \& {Prakash}, M. 2004, Sci, 304, 536

\bibitem[{{Lattimer} \& {Prakash}(2007)}]{LP07}
{Lattimer}, J.~M. \& {Prakash}, M. 2007, \physrep, 442, 109

\bibitem[{{Lewin} {et~al.}(1993){Lewin}, {van Paradijs}, \& {Taam}}]{LvPT93}
{Lewin}, W.~H.~G., {van Paradijs}, J., \& {Taam}, R.~E. 1993, Space Science
  Reviews, 62, 223

\bibitem[{{London} {et~al.}(1984){London}, {Howard}, \& {Taam}}]{London84}
{London}, R.~A., {Howard}, W.~M., \& {Taam}, R.~E. 1984, \apjl, 287, L27

\bibitem[{{London} {et~al.}(1986){London}, {Taam}, \& {Howard}}]{London86}
{London}, R.~A., {Taam}, R.~E., \& {Howard}, W.~M. 1986, \apj, 306, 170

\bibitem[{{Majczyna} {et~al.}(2005){Majczyna}, {Madej}, {Joss}, \&
  {R{\'o}{\.z}a{\'n}ska}}]{MMJR05}
{Majczyna}, A., {Madej}, J., {Joss}, P.~C., \& {R{\'o}{\.z}a{\'n}ska}, A. 2005,
  \aap, 430, 643

\bibitem[{{Nagirner} \& {Poutanen}(1994)}]{NP94a}
{Nagirner}, D.~I. \& {Poutanen}, J. 1994, Astrophysics and Space Physics
  Reviews, 9, 1

\bibitem[{{Olson} \& {Kunasz}(1987)}]{OK87}
{Olson}, G.~L. \& {Kunasz}, P.~B. 1987, \jqsrt, 38, 325

\bibitem[{{Papitto} {et~al.}(2011){Papitto}, {D'A{\`i}}, {Motta}, {Riggio},
  {Burderi}, {di Salvo}, {Belloni}, \& {Iaria}}]{pap11}
{Papitto}, A., {D'A{\`i}}, A., {Motta}, S., {et~al.} 2011, \aap, 526, L3

\bibitem[{{Parikh} {et~al.}(2013){Parikh}, {Jos{\'e}}, {Sala}, \&
  {Iliadis}}]{PJSI13}
{Parikh}, A., {Jos{\'e}}, J., {Sala}, G., \& {Iliadis}, C. 2013, Progress in
  Particle and Nuclear Physics, 69, 225

\bibitem[{{Pavlov} {et~al.}(1991){Pavlov}, {Shibanov}, \& {Zavlin}}]{PSZ91}
{Pavlov}, G.~G., {Shibanov}, I.~A., \& {Zavlin}, V.~E. 1991, \mnras, 253, 193

\bibitem[{{Penninx} {et~al.}(1989){Penninx}, {Damen}, {van Paradijs}, {Tan}, \&
  {Lewin}}]{Penninx89}
{Penninx}, W., {Damen}, E., {van Paradijs}, J., {Tan}, J., \& {Lewin}, W.~H.~G.
  1989, \aap, 208, 146

\bibitem[{{Poutanen} {et~al.}(2014){Poutanen}, {N{\"a}ttil{\"a}}, {Kajava},
  {Latvala}, {Galloway}, {Kuulkers}, \& {Suleimanov}}]{PNK14}
{Poutanen}, J., {N{\"a}ttil{\"a}}, J., {Kajava}, J.~J.~E., {et~al.} 2014,
  \mnras, 442, 3777

\bibitem[{{Poutanen} \& {Svensson}(1996)}]{PS96}
{Poutanen}, J. \& {Svensson}, R. 1996, \apj, 470, 249

\bibitem[{{Prasad} {et~al.}(1986){Prasad}, {Kershaw}, \& {Beason}}]{PKB86}
{Prasad}, M.~K., {Kershaw}, D.~S., \& {Beason}, J.~D. 1986, Appl. Phys. Lett.,
  48, 1193

\bibitem[{{Rauch} {et~al.}(2008){Rauch}, {Suleimanov}, \& {Werner}}]{RSW08}
{Rauch}, T., {Suleimanov}, V., \& {Werner}, W. 2008, \aap, 490, 1127
  
\bibitem[{{Rogers} {et~al.}(1996){Rogers}, {Swenson}, \& {Iglesias}}]{opal96}
{Rogers}, F.~J., {Swenson}, F.~J., \& {Iglesias}, C.~A. 1996, \apj, 456, 902

\bibitem[{{Schatz}(2011)}]{Schatz11}
{Schatz}, H. 2011, Progress in Particle and Nuclear Physics, 66, 277

\bibitem[{{Seaton} {et~al.}(1994){Seaton}, {Yan}, {Mihalas}, \&
  {Pradhan}}]{Seaton94}
{Seaton}, M.~J., {Yan}, Y., {Mihalas}, D., \& {Pradhan}, A.~K. 1994, \mnras,
  266, 805

\bibitem[{{Strohmayer} \& {Bildsten}(2006)}]{SB06}
{Strohmayer}, T. \& {Bildsten}, L. 2006, in Compact stellar X-ray sources,
  Cambridge Astrophysics Series, No. 39, ed. W.~{Lewin} \& M.~{van der Klis}
  (Cambridge: Cambridge University Press), 113--156

\bibitem[{{Strohmayer} \& {Brown}(2002)}]{SB02}
{Strohmayer}, T.~E. \& {Brown}, E.~F. 2002, \apj, 566, 1045

\bibitem[{{Suleimanov} {et~al.}(2014){Suleimanov}, {Klochkov}, {Pavlov}, \&
    {Werner}}]{SKP14}
{Suleimanov}, V., {Klochkov}, D., {Pavlov}, G., \& {Werner}, K. 2014, \apjs,
  210, 13
    
\bibitem[{{Suleimanov} \& {Poutanen}(2006)}]{SulP06}
{Suleimanov}, V. \& {Poutanen}, J. 2006, \mnras, 369, 2036

\bibitem[{{Suleimanov} {et~al.}(2011{\natexlab{a}}){Suleimanov}, {Poutanen},
  {Revnivtsev}, \& {Werner}}]{SPRW11}
{Suleimanov}, V., {Poutanen}, J., {Revnivtsev}, M., \& {Werner}, K.
  2011{\natexlab{a}}, \apj, 742, 122

\bibitem[{{Suleimanov} {et~al.}(2011{\natexlab{b}}){Suleimanov}, {Poutanen}, \&
  {Werner}}]{SPW11}
{Suleimanov}, V., {Poutanen}, J., \& {Werner}, K. 2011{\natexlab{b}}, \aap,
  527, A139

\bibitem[{{Suleimanov} {et~al.}(2012){Suleimanov}, {Poutanen}, \&
  {Werner}}]{SPW12}
{Suleimanov}, V., {Poutanen}, J., \& {Werner}, K. 2012, \aap, 545, A120

\bibitem[{{Suleimanov} \& {Werner}(2007)}]{SW07}
{Suleimanov}, V. \& {Werner}, K. 2007, \aap, 466, 661

\bibitem[{{Sutherland}(1998)}]{Su98}
{Sutherland}, R.~S. 1998, \mnras, 300, 321

\bibitem[{{van Paradijs} {et~al.}(1990){van Paradijs}, {Dotani}, {Tanaka}, \&
  {Tsuru}}]{vP90}
{van Paradijs}, J., {Dotani}, T., {Tanaka}, Y., \& {Tsuru}, T. 1990, \pasj, 42,
  633

\bibitem[{{Verner} {et~al.}(1996){Verner}, {Ferland}, {Korista}, \&
  {Yakovlev}}]{V96}
{Verner}, D.~A., {Ferland}, G.~J., {Korista}, K.~T., \& {Yakovlev}, D.~G. 1996,
  \apj, 465, 487

\bibitem[{{Verner} \& {Yakovlev}(1995)}]{VY95}
{Verner}, D.~A. \& {Yakovlev}, D.~G. 1995, \aaps, 109, 125

\bibitem[{{Weinberg} {et~al.}(2006){Weinberg}, {Bildsten}, \& {Schatz}}]{WBS06}
{Weinberg}, N.~N., {Bildsten}, L., \& {Schatz}, H. 2006, \apj, 639, 1018

\bibitem[{{Woosley} {et~al.}(2004){Woosley}, {Heger}, {Cumming}, {Hoffman},
  {Pruet}, {Rauscher}, {Fisker}, {Schatz}, {Brown}, \& {Wiescher}}]{Woosley04}
{Woosley}, S.~E., {Heger}, A., {Cumming}, A., {et~al.} 2004, \apjs, 151, 75

\bibitem[{{Woosley} \& {Taam}(1976)}]{WT76}
{Woosley}, S.~E. \& {Taam}, R.~E. 1976, \nat, 263, 101

\bibitem[{{Zhang} {et~al.}(2011){Zhang}, {M{\'e}ndez}, \& {Altamirano}}]{ZMA11}
{Zhang}, G., {M{\'e}ndez}, M., \& {Altamirano}, D. 2011, \mnras, 413, 1913

\end{thebibliography}

\begin{appendix}

\section{Validity of ideal plasma assumption}\label{sect:appendixA}

\begin{figure}
\centering
\includegraphics[width=8.5cm]{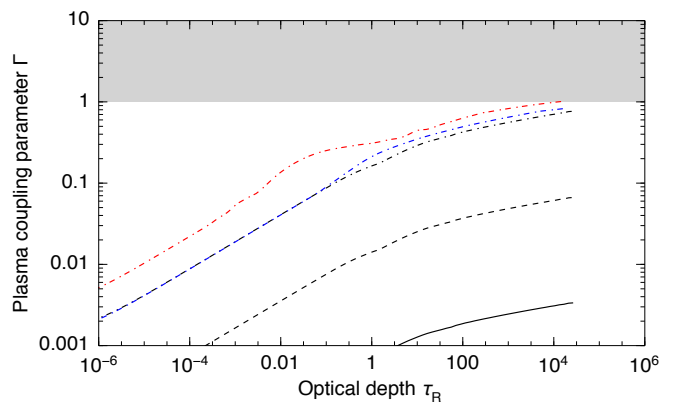}
\caption{\label{fig:gamma}
Plasma coupling parameter $\Gamma$ as a function of Rosseland mean optical depth $\tau_{\mathrm{R}}$.
Gray area marks the regime of (strong) coupling where our original assumption of ideal gas law slowly breaks.
Presented are coupling parameters for atmosphere compositions of pure iron (red lines) and solar mixture of H and He with metallicity $Z = 1Z_{\odot}$ (SolA1, black lines) and $Z = 40 Z_{\odot}$ (SolA40, blue lines) computed for hydrogen (solid lines), carbon (dashed) and iron (dot-dashed) ions.
}
\end{figure}

Compared to previous atmosphere models computed with hydrogen or helium compositions our work differs substantially due to high number densities of metals producing large electron number densities.
In the very deep layers, the dense plasma might become strongly coupled invalidating our assumption of ideal gas law \eqref{eq:idealgas}.
To quantify the depth where the coupling starts to be important, we can compare the ion specific Coulomb energies $E_{\mathrm{C}}$ and local thermal energies $E_{\mathrm{Th}}$ to get the so-called plasma coupling parameter, which is defined as
\be
\Gamma = \frac{E_{\mathrm{C}}}{E_{\mathrm{Th}}} = \frac{\Zc^2 e^2}{r_{\mathrm{s}}kT},
\ee
where $r_{\mathrm{s}} = \left( \frac{3}{4\pi n_{\mathrm{e}}} \right)^{1/3}$ is the Wigner-Seitz radius, i.e. the typical inter-particle distance, $\Zc$ is the charge of the ion and $n_{\mathrm{e}}$ is the local electron density \citep[see e.g.][for basic plasma parameters]{Chen84}.
For an ideal plasma (with equation of state that of the ideal gas), the thermal (kinetic) energy of ions is larger than the interaction potentials.
This corresponds to the situation $\Gamma \ll 1$.
In the opposite regime, pair distribution of ions becomes strongly localized and the plasma starts to crystallize into an (ideal) solid state.
Between these two states there exists a mixed phase of fluid-like plasma around $\Gamma \sim 1$.
The area where our approach is valid is limited to the case of the weakly coupled plasma, $\Gamma \ll 1$. 
 
The $\Gamma$-values are presented in Fig. \ref{fig:gamma} for different compositions and ions. 
The regime of intermediate coupling $\Gamma \sim 1$ is reached only for heavier ions in the electron-rich dense layers 
at Rosseland optical depths $\gtrsim10^4$. 
Thus this cannot affect the emergent spectra.
However, this can cause problems if the maximum column depth $m_{\mathrm{max}}$ is chosen to be too large so that the regime $\Gamma \gtrsim 1$ is reached within our computational domain.
In these very deep layers the inter-particle effects may become substantial, leading to a numerically challenging instability between the strong (and unphysical) recombination via the Saha equation and forced ionization because of the occupation probability formalism used.
This is a known problem because Saha's model assumes ideal gas conditions (i.e. every effect of the plasma is asserted on its ions and atoms internal structures) and uses only the ground state configurations in the ionization balance equations \citep{opal96}.
It is also worth mentioning that the pressure ionization method used here is a more of a phenomenological correction scheme as its corrections to non-ideal effects are obtained by minimizing a free energy of predefined gas \citep{HM88}.
More fundamental and physical approach would be the (activity) expansion of the grand canonical partition function of the plasma where pressure ionization is a consequence of the theory \citep[see e.g.][]{opal96}.
Such a fundamental treatment of plasma is, however, extremely challenging and out of the scope of this paper.

\section{Color correction tables}\label{sect:appB}

\topcaption{Color correction and dilution factors from the blackbody fits.}
 
\tablehead{$\ell$ & $g_{\mathrm{rad}}/g$ & $T_\mathrm{eff}$ (keV) & $f_{\mathrm{c}}$ & $w$ & $w f_{\mathrm{c}}^{4}$ \\
  \hline
  \hline
}

\begin{xtabular}[c]{c c c c c c}
  \multicolumn{6}{c}{$\quad$ $Z= Z_{\odot}$ $\quad$ $\log g = 14.0$ $\quad$ $T_{\mathrm{Edd}}= 1.70$ keV} \\
  \multicolumn{6}{c}{$R=14.80$ km $\quad$ $z=0.18$} \\
\hline
0.001 & 0.004 & 0.303 & 2.292 & 0.005 & 0.124 \\
0.003 & 0.007 & 0.398 & 1.843 & 0.024 & 0.275 \\
0.010 & 0.015 & 0.538 & 1.494 & 0.124 & 0.616 \\
0.030 & 0.035 & 0.708 & 1.383 & 0.240 & 0.878 \\
0.050 & 0.054 & 0.805 & 1.363 & 0.274 & 0.946 \\
0.070 & 0.074 & 0.875 & 1.346 & 0.298 & 0.978 \\
0.100 & 0.103 & 0.957 & 1.327 & 0.321 & 0.994 \\
0.150 & 0.152 & 1.059 & 1.324 & 0.321 & 0.986 \\
0.200 & 0.120 & 1.138 & 1.340 & 0.301 & 0.972 \\
0.300 & 0.294 & 1.259 & 1.383 & 0.263 & 0.962 \\
0.400 & 0.385 & 1.353 & 1.420 & 0.237 & 0.961 \\
0.500 & 0.478 & 1.431 & 1.452 & 0.218 & 0.971 \\
0.550 & 0.520 & 1.465 & 1.464 & 0.211 & 0.969 \\
0.600 & 0.568 & 1.498 & 1.478 & 0.204 & 0.973 \\
0.650 & 0.614 & 1.528 & 1.492 & 0.197 & 0.975 \\
0.700 & 0.660 & 1.557 & 1.507 & 0.190 & 0.977 \\
0.750 & 0.707 & 1.584 & 1.525 & 0.182 & 0.983 \\
0.800 & 0.753 & 1.609 & 1.548 & 0.172 & 0.988 \\
0.850 & 0.799 & 1.634 & 1.571 & 0.162 & 0.988 \\
0.900 & 0.844 & 1.657 & 1.601 & 0.151 & 0.993 \\
0.940 & 0.880 & 1.676 & 1.636 & 0.139 & 0.999 \\
0.960 & 0.898 & 1.684 & 1.657 & 0.133 & 1.000 \\
0.980 & 0.916 & 1.693 & 1.673 & 0.128 & 1.001 \\
1.000 & 0.933 & 1.702 & 1.701 & 0.120 & 1.004 \\
1.020 & 0.951 & 1.710 & 1.738 & 0.111 & 1.008 \\
1.040 & 0.967 & 1.718 & 1.781 & 0.101 & 1.012 \\
1.060 & 0.985 & 1.727 & 1.821 & 0.092 & 1.013 \\
\hline
  \multicolumn{6}{c}{$\quad$ $Z= Z_{\odot}$ $\quad$ $\log g = 14.3$ $\quad$ $T_{\mathrm{Edd}}= 2.02$ keV} \\
  \multicolumn{6}{c}{$R=10.88$ km $\quad$ $z=0.27$} \\
\hline
0.001 & 0.004 & 0.360 & 2.130 & 0.012 & 0.241 \\
0.003 & 0.006 & 0.473 & 1.741 & 0.048 & 0.444 \\
0.010 & 0.013 & 0.639 & 1.462 & 0.165 & 0.756 \\
0.030 & 0.032 & 0.842 & 1.350 & 0.286 & 0.948 \\
0.050 & 0.052 & 0.956 & 1.308 & 0.339 & 0.993 \\
0.070 & 0.071 & 1.040 & 1.287 & 0.364 & 0.999 \\
0.100 & 0.101 & 1.137 & 1.284 & 0.362 & 0.986 \\
0.150 & 0.148 & 1.259 & 1.310 & 0.328 & 0.964 \\
0.200 & 0.195 & 1.353 & 1.340 & 0.296 & 0.955 \\
0.300 & 0.287 & 1.497 & 1.388 & 0.257 & 0.955 \\
0.400 & 0.380 & 1.608 & 1.423 & 0.236 & 0.967 \\
0.500 & 0.471 & 1.701 & 1.449 & 0.221 & 0.977 \\
0.550 & 0.517 & 1.742 & 1.462 & 0.215 & 0.980 \\
0.600 & 0.564 & 1.780 & 1.473 & 0.208 & 0.979 \\
0.650 & 0.610 & 1.816 & 1.491 & 0.201 & 0.992 \\
0.700 & 0.655 & 1.850 & 1.507 & 0.193 & 0.996 \\
0.750 & 0.701 & 1.882 & 1.522 & 0.185 & 0.992 \\
0.800 & 0.739 & 1.913 & 1.535 & 0.176 & 0.978 \\
0.850 & 0.789 & 1.942 & 1.564 & 0.165 & 0.987 \\
0.900 & 0.829 & 1.970 & 1.592 & 0.153 & 0.984 \\
0.940 & 0.866 & 1.991 & 1.624 & 0.142 & 0.987 \\
0.960 & 0.885 & 2.002 & 1.644 & 0.136 & 0.992 \\
0.980 & 0.901 & 2.012 & 1.667 & 0.129 & 0.992 \\
1.000 & 0.915 & 2.022 & 1.687 & 0.122 & 0.990 \\
1.020 & 0.931 & 2.033 & 1.720 & 0.113 & 0.991 \\
1.040 & 0.947 & 2.042 & 1.757 & 0.104 & 0.992 \\
1.060 & 0.968 & 2.052 & 1.814 & 0.092 & 0.999 \\
1.080 & 0.988 & 2.062 & 1.871 & 0.081 & 0.987 \\
\hline
\multicolumn{6}{c}{$\quad$ $Z= Z_{\odot}$ $\quad$ $\log g = 14.6$ $\quad$ $T_{\mathrm{Edd}}= 2.40$ keV} \\
\multicolumn{6}{c}{$R=8.16$ km $\quad$ $z=0.42$} \\
\hline
0.001 & 0.003 & 0.427 & 2.052 & 0.021 & 0.365 \\
0.003 & 0.006 & 0.563 & 1.705 & 0.069 & 0.581 \\
0.010 & 0.013 & 0.760 & 1.433 & 0.203 & 0.857 \\
0.030 & 0.031 & 1.000 & 1.284 & 0.369 & 1.003 \\
0.050 & 0.049 & 1.137 & 1.242 & 0.424 & 1.007 \\
0.070 & 0.067 & 1.236 & 1.241 & 0.415 & 0.986 \\
0.100 & 0.095 & 1.352 & 1.266 & 0.374 & 0.961 \\
0.150 & 0.141 & 1.496 & 1.314 & 0.318 & 0.948 \\
0.200 & 0.188 & 1.607 & 1.351 & 0.285 & 0.950 \\
0.300 & 0.281 & 1.779 & 1.396 & 0.254 & 0.964 \\
0.400 & 0.372 & 1.912 & 1.420 & 0.239 & 0.970 \\
0.500 & 0.463 & 2.021 & 1.441 & 0.227 & 0.977 \\
0.550 & 0.509 & 2.070 & 1.454 & 0.220 & 0.982 \\
0.600 & 0.555 & 2.115 & 1.465 & 0.213 & 0.982 \\
0.650 & 0.600 & 2.158 & 1.482 & 0.205 & 0.992 \\
0.700 & 0.644 & 2.199 & 1.500 & 0.197 & 0.997 \\
0.750 & 0.688 & 2.237 & 1.517 & 0.188 & 0.998 \\
0.800 & 0.731 & 2.273 & 1.535 & 0.179 & 0.993 \\
0.850 & 0.774 & 2.308 & 1.553 & 0.169 & 0.982 \\
0.900 & 0.819 & 2.341 & 1.583 & 0.157 & 0.987 \\
0.940 & 0.857 & 2.367 & 1.616 & 0.145 & 0.991 \\
0.960 & 0.874 & 2.379 & 1.633 & 0.140 & 0.991 \\
0.980 & 0.890 & 2.392 & 1.652 & 0.133 & 0.992 \\
1.000 & 0.909 & 2.404 & 1.677 & 0.126 & 0.997 \\
1.020 & 0.925 & 2.416 & 1.703 & 0.119 & 0.998 \\
1.040 & 0.938 & 2.427 & 1.732 & 0.111 & 0.996 \\
1.060 & 0.953 & 2.439 & 1.774 & 0.101 & 0.998 \\
1.080 & 0.970 & 2.450 & 1.834 & 0.089 & 1.003 \\
1.100 & 0.986 & 2.462 & 1.894 & 0.076 & 0.985 \\
\hline
  \multicolumn{6}{c}{$\quad$ $Z= 20Z_{\odot}$ $\quad$ $\log g = 14.0$ $\quad$ $T_{\mathrm{Edd}}= 1.75$ keV} \\
  \multicolumn{6}{c}{$R=14.80$ km $\quad$ $z=0.18$} \\
\hline
0.001 & 0.017 & 0.311 & 1.805 & 0.002 & 0.022 \\
0.003 & 0.026 & 0.409 & 1.508 & 0.023 & 0.121 \\ 
0.010 & 0.043 & 0.553 & 1.286 & 0.185 & 0.505 \\ 
0.030 & 0.067 & 0.728 & 1.229 & 0.371 & 0.847 \\ 
0.050 & 0.087 & 0.827 & 1.233 & 0.401 & 0.927 \\ 
0.070 & 0.106 & 0.900 & 1.235 & 0.416 & 0.969 \\ 
0.100 & 0.139 & 0.984 & 1.232 & 0.438 & 1.009 \\ 
0.150 & 0.190 & 1.089 & 1.214 & 0.484 & 1.049 \\ 
0.200 & 0.240 & 1.170 & 1.192 & 0.529 & 1.066 \\ 
0.300 & 0.332 & 1.295 & 1.177 & 0.546 & 1.046 \\ 
0.400 & 0.413 & 1.391 & 1.201 & 0.496 & 1.033 \\ 
0.500 & 0.526 & 1.471 & 1.238 & 0.415 & 0.975 \\ 
0.550 & 0.568 & 1.507 & 1.258 & 0.386 & 0.968 \\ 
0.600 & 0.609 & 1.540 & 1.280 & 0.359 & 0.964 \\ 
0.650 & 0.651 & 1.571 & 1.302 & 0.335 & 0.962 \\ 
0.700 & 0.692 & 1.600 & 1.325 & 0.312 & 0.962 \\ 
0.750 & 0.734 & 1.628 & 1.351 & 0.290 & 0.966 \\ 
0.800 & 0.775 & 1.655 & 1.378 & 0.270 & 0.972 \\ 
0.850 & 0.816 & 1.680 & 1.404 & 0.251 & 0.975 \\ 
0.900 & 0.857 & 1.704 & 1.435 & 0.232 & 0.981 \\ 
0.940 & 0.890 & 1.723 & 1.464 & 0.216 & 0.990 \\ 
0.960 & 0.907 & 1.732 & 1.478 & 0.207 & 0.990 \\ 
0.980 & 0.921 & 1.741 & 1.494 & 0.199 & 0.993 \\ 
1.000 & 0.939 & 1.750 & 1.519 & 0.188 & 1.000 \\ 
1.020 & 0.953 & 1.758 & 1.551 & 0.172 & 0.998 \\ 
1.040 & 0.972 & 1.767 & 1.599 & 0.154 & 1.008 \\ 
1.060 & 0.989 & 1.775 & 1.640 & 0.140 & 1.012 \\ 
\hline
\multicolumn{6}{c}{$\quad$ $Z= 20Z_{\odot}$ $\quad$ $\log g = 14.3$ $\quad$ $T_{\mathrm{Edd}}= 2.08$ keV} \\
\multicolumn{6}{c}{$R=10.88$ km $\quad$ $z=0.27$} \\
\hline
0.001 & 0.015 & 0.370 & 1.872 & 0.005 & 0.063 \\ 
0.003 & 0.024 & 0.487 & 1.521 & 0.049 & 0.262 \\ 
0.010 & 0.037 & 0.657 & 1.269 & 0.277 & 0.720 \\ 
0.030 & 0.059 & 0.865 & 1.240 & 0.397 & 0.940 \\ 
0.050 & 0.080 & 0.983 & 1.228 & 0.442 & 1.005 \\ 
0.070 & 0.103 & 1.069 & 1.211 & 0.486 & 1.043 \\ 
0.100 & 0.136 & 1.169 & 1.181 & 0.553 & 1.077 \\ 
0.150 & 0.186 & 1.294 & 1.145 & 0.630 & 1.083 \\ 
0.200 & 0.230 & 1.391 & 1.135 & 0.638 & 1.056 \\ 
0.300 & 0.312 & 1.539 & 1.147 & 0.576 & 0.998 \\ 
0.400 & 0.390 & 1.654 & 1.182 & 0.493 & 0.964 \\ 
0.500 & 0.471 & 1.748 & 1.226 & 0.420 & 0.951 \\ 
0.550 & 0.510 & 1.791 & 1.248 & 0.391 & 0.949 \\ 
0.600 & 0.548 & 1.830 & 1.269 & 0.366 & 0.950 \\ 
0.650 & 0.589 & 1.867 & 1.292 & 0.342 & 0.952 \\ 
0.700 & 0.633 & 1.902 & 1.315 & 0.320 & 0.956 \\ 
0.750 & 0.704 & 1.935 & 1.346 & 0.290 & 0.952 \\ 
0.800 & 0.757 & 1.966 & 1.384 & 0.265 & 0.973 \\ 
0.850 & 0.805 & 1.996 & 1.410 & 0.248 & 0.980 \\ 
0.900 & 0.845 & 2.025 & 1.433 & 0.233 & 0.984 \\ 
0.940 & 0.875 & 2.047 & 1.455 & 0.220 & 0.986 \\ 
0.960 & 0.891 & 2.058 & 1.470 & 0.212 & 0.989 \\ 
0.980 & 0.907 & 2.069 & 1.486 & 0.203 & 0.991 \\ 
1.000 & 0.930 & 2.079 & 1.512 & 0.193 & 1.007 \\ 
1.020 & 0.942 & 2.090 & 1.537 & 0.179 & 0.998 \\ 
1.040 & 0.953 & 2.100 & 1.576 & 0.161 & 0.994 \\ 
1.060 & 0.970 & 2.110 & 1.639 & 0.139 & 1.000 \\ 
1.080 & 0.976 & 2.120 & 1.730 & 0.111 & 0.997 \\ 
\hline
\multicolumn{6}{c}{$\quad$ $Z= 20Z_{\odot}$ $\quad$ $\log g = 14.6$ $\quad$ $T_{\mathrm{Edd}}= 2.47$ keV} \\
\multicolumn{6}{c}{$R=8.16$ km $\quad$ $z=0.42$} \\
\hline
0.001 & 0.014 & 0.439 & 2.077 & 0.007 & 0.123 \\ 
0.003 & 0.021 & 0.578 & 1.616 & 0.057 & 0.391 \\ 
0.010 & 0.032 & 0.781 & 1.310 & 0.280 & 0.824 \\ 
0.030 & 0.054 & 1.028 & 1.220 & 0.466 & 1.031 \\ 
0.050 & 0.077 & 1.169 & 1.170 & 0.584 & 1.093 \\ 
0.070 & 0.099 & 1.271 & 1.130 & 0.685 & 1.118 \\ 
0.100 & 0.130 & 1.390 & 1.097 & 0.765 & 1.107 \\ 
0.150 & 0.176 & 1.538 & 1.089 & 0.746 & 1.048 \\ 
0.200 & 0.217 & 1.653 & 1.103 & 0.675 & 0.998 \\ 
0.300 & 0.296 & 1.829 & 1.152 & 0.538 & 0.949 \\ 
0.400 & 0.372 & 1.965 & 1.205 & 0.444 & 0.937 \\ 
0.500 & 0.446 & 2.078 & 1.251 & 0.384 & 0.941 \\ 
0.550 & 0.483 & 2.128 & 1.272 & 0.361 & 0.944 \\ 
0.600 & 0.520 & 2.175 & 1.290 & 0.343 & 0.949 \\ 
0.650 & 0.558 & 2.219 & 1.307 & 0.327 & 0.954 \\ 
0.700 & 0.597 & 2.260 & 1.324 & 0.312 & 0.960 \\ 
0.750 & 0.638 & 2.300 & 1.341 & 0.299 & 0.965 \\ 
0.800 & 0.680 & 2.337 & 1.357 & 0.286 & 0.971 \\ 
0.850 & 0.720 & 2.373 & 1.372 & 0.275 & 0.975 \\ 
0.900 & 0.761 & 2.407 & 1.390 & 0.265 & 0.987 \\ 
0.940 & 0.786 & 2.433 & 1.397 & 0.257 & 0.978 \\ 
0.960 & 0.806 & 2.446 & 1.406 & 0.252 & 0.987 \\ 
0.980 & 0.822 & 2.459 & 1.414 & 0.248 & 0.991 \\ 
1.000 & 0.837 & 2.471 & 1.422 & 0.242 & 0.991 \\ 
1.020 & 0.855 & 2.483 & 1.434 & 0.235 & 0.991 \\ 
1.040 & 0.873 & 2.495 & 1.447 & 0.226 & 0.989 \\ 
1.060 & 0.894 & 2.507 & 1.466 & 0.216 & 0.995 \\ 
1.080 & 0.952 & 2.519 & 1.515 & 0.187 & 0.986 \\ 
1.100 & 0.998 & 2.533 & 1.565 & 0.158 & 0.949 \\ 
\hline
\multicolumn{6}{c}{$\quad$ $Z= 40Z_{\odot}$ $\quad$ $\log g = 14.0$ $\quad$ $T_{\mathrm{Edd}}= 1.81$ keV} \\
\multicolumn{6}{c}{$R=14.80$ km $\quad$ $z=0.18$} \\
\hline
0.001 & 0.026 & 0.322 & 1.503 & 0.006 & 0.032 \\ 
0.003 & 0.041 & 0.424 & 1.335 & 0.048 & 0.151 \\ 
0.010 & 0.064 & 0.573 & 1.228 & 0.239 & 0.544 \\ 
0.030 & 0.093 & 0.754 & 1.213 & 0.394 & 0.851 \\ 
0.050 & 0.113 & 0.857 & 1.222 & 0.415 & 0.926 \\ 
0.070 & 0.131 & 0.932 & 1.226 & 0.429 & 0.967 \\ 
0.100 & 0.163 & 1.019 & 1.223 & 0.451 & 1.008 \\ 
0.150 & 0.214 & 1.128 & 1.206 & 0.498 & 1.053 \\ 
0.200 & 0.263 & 1.212 & 1.180 & 0.557 & 1.079 \\ 
0.300 & 0.354 & 1.341 & 1.141 & 0.640 & 1.084 \\ 
0.400 & 0.428 & 1.441 & 1.126 & 0.654 & 1.053 \\ 
0.500 & 0.505 & 1.524 & 1.133 & 0.614 & 1.012 \\ 
0.550 & 0.543 & 1.561 & 1.143 & 0.585 & 0.999 \\ 
0.600 & 0.599 & 1.595 & 1.168 & 0.531 & 0.988 \\ 
0.650 & 0.671 & 1.627 & 1.208 & 0.451 & 0.960 \\ 
0.700 & 0.726 & 1.657 & 1.241 & 0.402 & 0.954 \\ 
0.750 & 0.763 & 1.686 & 1.266 & 0.371 & 0.953 \\ 
0.800 & 0.800 & 1.714 & 1.293 & 0.341 & 0.954 \\ 
0.850 & 0.837 & 1.740 & 1.323 & 0.312 & 0.957 \\ 
0.900 & 0.873 & 1.765 & 1.357 & 0.284 & 0.963 \\ 
0.940 & 0.900 & 1.784 & 1.387 & 0.261 & 0.967 \\ 
0.960 & 0.913 & 1.794 & 1.404 & 0.249 & 0.970 \\ 
0.980 & 0.931 & 1.803 & 1.426 & 0.237 & 0.979 \\ 
1.000 & 0.943 & 1.812 & 1.447 & 0.224 & 0.980 \\ 
1.020 & 0.960 & 1.821 & 1.476 & 0.209 & 0.994 \\ 
1.040 & 0.974 & 1.830 & 1.513 & 0.191 & 0.999 \\ 
1.060 & 0.989 & 1.839 & 1.559 & 0.170 & 1.004 \\ 
\hline
\multicolumn{6}{c}{$\quad$ $Z= 40Z_{\odot}$ $\quad$ $\log g = 14.3$ $\quad$ $T_{\mathrm{Edd}}= 2.15$ keV} \\
\multicolumn{6}{c}{$R=10.88$ km $\quad$ $z=0.27$} \\
\hline
0.001 & 0.024 & 0.383 & 1.621 & 0.010 & 0.071 \\ 
0.003 & 0.036 & 0.504 & 1.362 & 0.091 & 0.313 \\ 
0.010 & 0.055 & 0.681 & 1.221 & 0.341 & 0.759 \\ 
0.030 & 0.080 & 0.896 & 1.231 & 0.412 & 0.945 \\ 
0.050 & 0.101 & 1.018 & 1.223 & 0.451 & 1.009 \\ 
0.070 & 0.126 & 1.108 & 1.208 & 0.493 & 1.049 \\ 
0.100 & 0.160 & 1.211 & 1.179 & 0.563 & 1.087 \\ 
0.150 & 0.209 & 1.340 & 1.131 & 0.679 & 1.112 \\ 
0.200 & 0.253 & 1.440 & 1.103 & 0.745 & 1.102 \\ 
0.300 & 0.327 & 1.594 & 1.082 & 0.762 & 1.046 \\ 
0.400 & 0.395 & 1.713 & 1.089 & 0.707 & 0.993 \\ 
0.500 & 0.465 & 1.811 & 1.114 & 0.622 & 0.958 \\ 
0.550 & 0.500 & 1.855 & 1.131 & 0.579 & 0.947 \\ 
0.600 & 0.533 & 1.895 & 1.147 & 0.542 & 0.940 \\ 
0.650 & 0.564 & 1.934 & 1.164 & 0.510 & 0.935 \\ 
0.700 & 0.595 & 1.970 & 1.181 & 0.480 & 0.933 \\ 
0.750 & 0.624 & 2.004 & 1.198 & 0.453 & 0.933 \\ 
0.800 & 0.654 & 2.037 & 1.215 & 0.429 & 0.934 \\ 
0.850 & 0.683 & 2.068 & 1.232 & 0.406 & 0.936 \\ 
0.900 & 0.724 & 2.098 & 1.256 & 0.375 & 0.935 \\ 
0.940 & 0.851 & 2.121 & 1.294 & 0.318 & 0.892 \\ 
0.960 & 0.860 & 2.132 & 1.333 & 0.293 & 0.925 \\ 
0.980 & 0.897 & 2.143 & 1.398 & 0.256 & 0.976 \\ 
1.000 & 0.924 & 2.154 & 1.431 & 0.233 & 0.977 \\ 
1.020 & 0.946 & 2.164 & 1.458 & 0.218 & 0.987 \\ 
1.040 & 0.964 & 2.175 & 1.490 & 0.202 & 0.996 \\ 
1.060 & 0.966 & 2.185 & 1.535 & 0.177 & 0.984 \\ 
1.080 & 0.971 & 2.196 & 1.623 & 0.141 & 0.982 \\ 
\hline
\multicolumn{6}{c}{$\quad$ $Z= 40Z_{\odot}$ $\quad$ $\log g = 14.6$ $\quad$ $T_{\mathrm{Edd}}= 2.56$ keV} \\
\multicolumn{6}{c}{$R=8.16$ km $\quad$ $z=0.42$} \\
\hline
0.001 & 0.022 & 0.455 & 1.859 & 0.011 & 0.125 \\ 
0.003 & 0.032 & 0.599 & 1.474 & 0.093 & 0.440 \\ 
0.010 & 0.048 & 0.809 & 1.272 & 0.327 & 0.855 \\ 
0.030 & 0.074 & 1.065 & 1.212 & 0.482 & 1.040 \\ 
0.050 & 0.103 & 1.210 & 1.165 & 0.599 & 1.104 \\ 
0.070 & 0.129 & 1.317 & 1.123 & 0.714 & 1.134 \\ 
0.100 & 0.164 & 1.439 & 1.080 & 0.835 & 1.135 \\ 
0.150 & 0.215 & 1.593 & 1.054 & 0.876 & 1.081 \\ 
0.200 & 0.252 & 1.712 & 1.066 & 0.833 & 1.074 \\ 
0.300 & 0.372 & 1.894 & 1.111 & 0.621 & 0.946 \\ 
0.400 & 0.455 & 2.036 & 1.162 & 0.509 & 0.928 \\ 
0.500 & 0.535 & 2.152 & 1.211 & 0.432 & 0.929 \\ 
0.550 & 0.575 & 2.204 & 1.233 & 0.403 & 0.932 \\ 
0.600 & 0.614 & 2.253 & 1.254 & 0.378 & 0.937 \\ 
0.650 & 0.653 & 2.298 & 1.275 & 0.357 & 0.942 \\ 
0.700 & 0.693 & 2.341 & 1.294 & 0.338 & 0.949 \\ 
0.750 & 0.732 & 2.382 & 1.313 & 0.322 & 0.955 \\ 
0.800 & 0.770 & 2.421 & 1.332 & 0.306 & 0.963 \\ 
0.850 & 0.809 & 2.458 & 1.350 & 0.292 & 0.968 \\ 
0.900 & 0.846 & 2.493 & 1.369 & 0.277 & 0.973 \\ 
0.940 & 0.877 & 2.520 & 1.388 & 0.264 & 0.979 \\ 
0.960 & 0.892 & 2.534 & 1.398 & 0.257 & 0.982 \\ 
0.980 & 0.905 & 2.547 & 1.409 & 0.249 & 0.982 \\ 
1.000 & 0.924 & 2.560 & 1.427 & 0.239 & 0.993 \\ 
1.020 & 0.938 & 2.572 & 1.444 & 0.228 & 0.992 \\ 
1.040 & 0.950 & 2.585 & 1.469 & 0.213 & 0.991 \\ 
1.060 & 0.958 & 2.597 & 1.504 & 0.193 & 0.986 \\ 
1.080 & 0.971 & 2.609 & 1.570 & 0.163 & 0.988 \\ 
1.100 & 0.981 & 2.621 & 1.665 & 0.129 & 0.991 \\ 
\hline
\multicolumn{6}{c}{$\quad$ Fe $\quad$ $\log g = 14.0$ $\quad$ $T_{\mathrm{Edd}}= 1.95$ keV} \\
\multicolumn{6}{c}{$R=14.80$ km $\quad$ $z=0.18$} \\
\hline
0.001 & 0.022 & 0.347 & 1.998 & 0.001 & 0.014 \\ 
0.003 & 0.047 & 0.457 & 2.061 & 0.007 & 0.130 \\ 
0.010 & 0.081 & 0.618 & 1.836 & 0.048 & 0.548 \\ 
0.030 & 0.182 & 0.813 & 1.587 & 0.137 & 0.866 \\ 
0.050 & 0.290 & 0.924 & 1.486 & 0.198 & 0.969 \\ 
0.070 & 0.370 & 1.005 & 1.425 & 0.248 & 1.024 \\ 
0.100 & 0.435 & 1.099 & 1.365 & 0.309 & 1.072 \\ 
0.150 & 0.521 & 1.216 & 1.304 & 0.385 & 1.114 \\ 
0.200 & 0.588 & 1.307 & 1.252 & 0.464 & 1.139 \\ 
0.300 & 0.692 & 1.446 & 1.148 & 0.667 & 1.160 \\ 
0.400 & 0.770 & 1.554 & 1.091 & 0.814 & 1.152 \\ 
0.500 & 0.833 & 1.643 & 1.060 & 0.880 & 1.112 \\ 
0.550 & 0.857 & 1.683 & 1.055 & 0.875 & 1.083 \\ 
0.600 & 0.880 & 1.720 & 1.055 & 0.850 & 1.054 \\ 
0.650 & 0.898 & 1.754 & 1.062 & 0.809 & 1.028 \\ 
0.700 & 0.915 & 1.787 & 1.072 & 0.759 & 1.003 \\ 
0.750 & 0.929 & 1.818 & 1.088 & 0.701 & 0.983 \\ 
0.800 & 0.946 & 1.848 & 1.108 & 0.641 & 0.967 \\ 
0.850 & 0.954 & 1.876 & 1.134 & 0.578 & 0.956 \\ 
0.900 & 0.970 & 1.903 & 1.165 & 0.514 & 0.948 \\ 
0.940 & 0.971 & 1.924 & 1.197 & 0.461 & 0.947 \\ 
0.960 & 0.982 & 1.934 & 1.214 & 0.436 & 0.947 \\ 
0.980 & 0.984 & 1.944 & 1.233 & 0.410 & 0.948 \\ 
1.000 & 0.987 & 1.954 & 1.254 & 0.385 & 0.950 \\ 
1.020 & 0.988 & 1.964 & 1.273 & 0.363 & 0.953 \\ 
1.040 & 0.989 & 1.973 & 1.298 & 0.338 & 0.959 \\ 
1.060 & 0.991 & 1.983 & 1.329 & 0.309 & 0.964 \\ 
\hline
\multicolumn{6}{c}{$\quad$ Fe $\quad$ $\log g = 14.3$ $\quad$ $T_{\mathrm{Edd}}= 2.32$ keV} \\
\multicolumn{6}{c}{$R=10.88$ km $\quad$ $z=0.27$} \\
\hline
0.001 & 0.025 & 0.413 & 2.117 & 0.004 & 0.072 \\ 
0.003 & 0.045 & 0.543 & 2.029 & 0.021 & 0.359 \\ 
0.010 & 0.074 & 0.734 & 1.727 & 0.085 & 0.753 \\ 
0.030 & 0.199 & 0.967 & 1.472 & 0.213 & 1.001 \\ 
0.050 & 0.307 & 1.098 & 1.369 & 0.307 & 1.080 \\ 
0.070 & 0.365 & 1.194 & 1.309 & 0.382 & 1.120 \\ 
0.100 & 0.428 & 1.306 & 1.253 & 0.468 & 1.154 \\ 
0.150 & 0.510 & 1.445 & 1.170 & 0.633 & 1.185 \\ 
0.200 & 0.576 & 1.553 & 1.098 & 0.820 & 1.192 \\ 
0.300 & 0.678 & 1.719 & 1.031 & 1.024 & 1.158 \\ 
0.400 & 0.757 & 1.847 & 1.006 & 1.060 & 1.084 \\ 
0.500 & 0.818 & 1.953 & 1.006 & 0.991 & 1.014 \\ 
0.550 & 0.843 & 2.000 & 1.013 & 0.936 & 0.986 \\ 
0.600 & 0.865 & 2.044 & 1.024 & 0.875 & 0.964 \\ 
0.650 & 0.885 & 2.085 & 1.040 & 0.809 & 0.946 \\ 
0.700 & 0.902 & 2.124 & 1.059 & 0.742 & 0.933 \\ 
0.750 & 0.918 & 2.161 & 1.082 & 0.675 & 0.925 \\ 
0.800 & 0.932 & 2.196 & 1.108 & 0.611 & 0.920 \\ 
0.850 & 0.945 & 2.230 & 1.138 & 0.549 & 0.920 \\ 
0.900 & 0.956 & 2.262 & 1.171 & 0.490 & 0.924 \\ 
0.940 & 0.964 & 2.287 & 1.201 & 0.446 & 0.929 \\ 
0.960 & 0.968 & 2.299 & 1.217 & 0.425 & 0.933 \\ 
0.980 & 0.972 & 2.311 & 1.234 & 0.404 & 0.937 \\ 
1.000 & 0.976 & 2.322 & 1.252 & 0.384 & 0.942 \\ 
1.020 & 0.979 & 2.334 & 1.270 & 0.364 & 0.947 \\ 
1.040 & 0.982 & 2.345 & 1.290 & 0.344 & 0.953 \\ 
1.060 & 0.985 & 2.356 & 1.312 & 0.324 & 0.960 \\ 
1.080 & 0.988 & 2.367 & 1.336 & 0.304 & 0.968 \\ 
\hline
\multicolumn{6}{c}{$\quad$ Fe $\quad$ $\log g = 14.6$ $\quad$ $T_{\mathrm{Edd}}= 2.76$ keV} \\
\multicolumn{6}{c}{$R=8.16$ km $\quad$ $z=0.42$} \\
\hline
0.001 & 0.025 & 0.491 & 2.080 & 0.013 & 0.238 \\ 
0.003 & 0.041 & 0.646 & 1.905 & 0.045 & 0.598 \\ 
0.010 & 0.068 & 0.873 & 1.573 & 0.152 & 0.929 \\ 
0.030 & 0.214 & 1.149 & 1.319 & 0.373 & 1.129 \\ 
0.050 & 0.306 & 1.305 & 1.223 & 0.530 & 1.187 \\ 
0.070 & 0.356 & 1.420 & 1.168 & 0.654 & 1.215 \\ 
0.100 & 0.416 & 1.552 & 1.090 & 0.875 & 1.236 \\ 
0.150 & 0.496 & 1.718 & 1.010 & 1.175 & 1.222 \\ 
0.200 & 0.560 & 1.846 & 0.975 & 1.297 & 1.171 \\ 
0.300 & 0.663 & 2.043 & 0.949 & 1.296 & 1.052 \\ 
0.400 & 0.740 & 2.195 & 0.956 & 1.157 & 0.965 \\ 
0.500 & 0.799 & 2.321 & 0.982 & 0.983 & 0.914 \\ 
0.550 & 0.824 & 2.377 & 1.001 & 0.896 & 0.899 \\ 
0.600 & 0.846 & 2.429 & 1.023 & 0.814 & 0.891 \\ 
0.650 & 0.866 & 2.478 & 1.047 & 0.737 & 0.887 \\ 
0.700 & 0.885 & 2.525 & 1.073 & 0.668 & 0.886 \\ 
0.750 & 0.902 & 2.569 & 1.101 & 0.605 & 0.890 \\ 
0.800 & 0.917 & 2.610 & 1.130 & 0.550 & 0.896 \\ 
0.850 & 0.931 & 2.650 & 1.159 & 0.500 & 0.904 \\ 
0.900 & 0.944 & 2.688 & 1.190 & 0.456 & 0.914 \\ 
0.940 & 0.954 & 2.718 & 1.214 & 0.424 & 0.923 \\ 
0.960 & 0.959 & 2.732 & 1.228 & 0.409 & 0.928 \\ 
0.980 & 0.963 & 2.746 & 1.241 & 0.394 & 0.934 \\ 
1.000 & 0.968 & 2.760 & 1.254 & 0.379 & 0.939 \\ 
1.020 & 0.972 & 2.774 & 1.268 & 0.365 & 0.945 \\ 
1.040 & 0.976 & 2.787 & 1.283 & 0.351 & 0.951 \\ 
1.060 & 0.980 & 2.801 & 1.299 & 0.336 & 0.958 \\ 
1.080 & 0.984 & 2.814 & 1.317 & 0.321 & 0.966 \\ 
1.100 & 0.987 & 2.827 & 1.336 & 0.306 & 0.974 \\
\hline
\end{xtabular}\label{tab:fc-w-l_A}
\clearpage
\end{appendix}

\end{document}